\documentstyle[11pt,aaspp4,epsf,psfig,epsfig]{article}

\newcommand\nion[2]{#1\,\lowercase{{\sc #2}}}

\newcommand\eps[1]{log~$\epsilon$(#1)}
\def\kmsec{\mbox{km~s$^{\rm -1}$}}
\def\teff{\mbox{T$_{\rm eff}$}}
\def\BmV0{\mbox{(B-V)$^{\rm o}$}}
\def\VmK0{\mbox{(V-K)$^{\rm o}$}}
\def\MV0{\mbox{M$_{\rm V}^{\rm o}$}}
\def\MV{\mbox{M$_{\rm V}$}}

\def\1st{{\it 1$^{st}$}}
\def\2nd{{\it 2$^{nd}$}}
\def\3rd{{\it 3$^{rd}$}}

\slugcomment{Accepted by ApJ}

\begin{document}

\title{Rubidium in Metal-Deficient Disk and Halo Stars}

\author{Jocelyn Tomkin and David L. Lambert}
\affil{Department of Astronomy and McDonald Observatory, University of 
Texas, Austin, TX 78712}
\authoremail{jt@alexis.as.utexas.edu, dll@astro.as.utexas.edu}

\begin{abstract}

We report the first extensive study of stellar Rb
abundances.  High-resolution spectra have been used to determine,
or set upper limits on, the abundances of this heavy element and
the associated elements Y, Zr, and Ba in 44 dwarfs and giants
with metallicities spanning the range $-2.0 <$ [Fe/H] $< 0.0$.
In metal-deficient stars Rb is systematically overabundant relative
to Fe; we find an average [Rb/Fe] of +0.21 for the 32 stars
with [Fe/H] $< -0.5$ and measured Rb.  This behavior contrasts
with that of Y, Zr, and Ba, which, with the exception of
three new CH stars (HD\,23439A and B and BD\,+5$^{\rm o}$\,3640), are
consistently slightly deficient relative to Fe in the same
stars; excluding the three CH stars, we find the stars with
[Fe/H] $< -0.5$ have average [Y/Fe], [Zr/Fe], and [Ba/Fe] of
--0.19 (24 stars), --0.12 (28 stars), and --0.06 (29 stars),
respectively.  The different behavior of Rb on the one hand
and Y, Zr, and Ba on the other can be attributed in part to the fact
that in the Sun and in these stars Rb has a large $r$-process
component while Y, Zr, and Ba are mostly $s$-process elements
with only small $r$-process components.
In addition, the Rb $s$-process abundance is dependent on the
neutron density at the $s$-processing site.  Published observations
of Rb in $s$-process enriched red giants indicate a higher neutron
density in the metal-poor giants.  These observations imply a higher
$s$-process abundance for Rb in metal-poor stars.  The calculated
combination of the Rb $r$-process abundance, as estimated for the
stellar Eu abundances, and the $s$-process abundance as estimated
for red giants accounts satisfactorily for the observed run of
[Rb/Fe] with [Fe/H].

\end{abstract}


\section{INTRODUCTION}

In the solar neighborhood reside stars of differing metallicity. Stars
with a metallicity of approximately the solar value belong overwhelmingly
to the Galactic disk. Stars of lower metallicity are on orbits that
identify them as members of the Galactic halo. The metallicity
[Fe/H] $\simeq -1$ may be taken as a rough boundary between disk and
halo stars. One tool for unravelling the evolution of the Galaxy is the
measurement of the chemical compositions of stars as a function of
metallicity. In this paper,
we present the first extensive series of measurements of the rubidium
abundance in disk and halo stars. Rubidium is potentially a special
diagnostic of the neutron capture $s$-process.

Rubidium is present in two isotopic forms: $^{85}$Rb which is stable and
$^{87}$Rb which with a half-life of 5 $\times 10^{10}$ yr may be deemed
effectively stable from the astrophysical point of view. As we remark
later, astronomical detection of Rb must rely on the Rb\,{\sc i}
lines that do not permit measurement of the relative isotopic Rb abundances
from stellar spectra. Hence, we discuss the elemental Rb abundance.
Analysis of the solar system abundances of Rb and adjoining elements
shows that the neutron capture $s$- and $r$-processes are about
equally responsible for the synthesis of Rb. Scrutiny of the $s$-process
abundances shows that the `main' $s$-process not the `weak' $s$-process
is the principal
source of Rb's $s$-process component. The main $s$-process which
manufactured elements heavier than about Rb is identified with the
He-burning shell of intermediate and low mass AGB stars. The weak $s$-process
is identified with He-core and possibly C-core burning of massive stars.

Evolution of the Galaxy's $s$ and $r$-process products is rather
directly observed from the stellar abundances of elements that
are predominantly attributable to either the $s$ or to the $r$-process.
Traditional tracers include Ba for the $s$-process and Eu for the
$r$-process.  Elucidation of the operation of the neutron capture processes
requires observations of more than a single element per process.
As an example, we note that the abundance ratio of a `light' to a `heavy'
$s$-process elements, say Zr to Ba, provides information on the
integrated exposure of material to neutrons. Rubidium with a roughly
equal mix of $s$ and $r$-process contributions, and an unfavorable
electronic structure for ready detection in stellar spectra would
seem to be an element of little interest.  Closer inspection of
the working of the $s$-process shows, however, that Rb offers a unique
insight into the process: Rb's role as a monitor of the neutron
density at the $s$-process site.

Along the $s$-process path, Rb is preceded
by krypton with the path entering Kr at $^{80}$Kr and exiting at either
$^{84}$Kr or $^{86}$Kr. Unstable $^{85}$Kr controls the exit. At low
neutron density at the $s$-process site, stable $^{84}$Kr is converted by
neutron capture to $^{85}$Kr that decays to stable $^{85}$Rb
with the path continuing to $^{86}$Sr. At high
neutron densities, $^{85}$Kr does not $\beta$-decay
but experiences a neutron capture and is converted
to stable $^{86}$Kr. Subsequent neutron
capture by $^{86}$Kr leads by $\beta$-decay of $^{87}$Kr to (effectively)
stable $^{87}$Rb. When a steady flow along the $s$-process path
is attained, the density of a nuclide is given approximately by
the condition that $\sigma_iN_i \simeq $constant where $\sigma_i$ is the
neutron capture cross-section of nuclide $i$ and $N_i$ is the abundance of
that nuclide. Since $\sigma_{87} \simeq \sigma_{85}/10$ for the Rb isotopes,
the switch of the $^{85}$Kr branch from its low neutron density routing
through $^{85}$Rb to its high neutron density routing through $^{87}$Rb
increases the total Rb abundance by about an order of magnitude relative
to the abundance of other elements in this section of the $s$-process
path, such as Sr, Y, and Zr. The isotopic mix of Rb is obviously altered
as a function of neutron density but this is not measureable for cool
stars.  (Krypton is undetectable spectroscopically in cool stars.)
Operation of the $^{85}$Kr branch is more complicated than sketched, for
example, neutron capture from $^{84}$Kr feeds not only the $^{85}$Kr
ground state but a short-lived isomeric state that at all reasonable
neutron densities provides some leakage to $^{85}$Rb. A thorough discussion
of the $^{85}$Kr branch is provided by Beer \& Macklin (1989)
\nocite{BM89} and its
use in determining the effective neutron density of the $s$-process in
stars is discussed by Tomkin \& Lambert (1983)\nocite{TL83}
and Lambert et al. (1995)\nocite{LSB95}.
When detailed abundance measurements are available as in the case of
the carbonaceous chondrites, several branches along the $s$-process
path serve as neutron densitometers but Rb is the only low neutron
density branch available to stellar spectroscopists. (At higher
neutron densities, a branch controlled by $^{95}$Zr is exploitable
in cool stars showing ZrO bands [Lambert et al. 1995\nocite{LSB95}].)

A primary reason for the near neglect of Rb in reports on quantitative
spectroscopy of stars is that it is a trace alkali element. The Rb atom's
low ionization potential (4.177\,eV) ensures that Rb is primarily ionized
but the rare-gas electronic structure of Rb$^+$ provides resonance
lines in the far ultraviolet. Detection of Rb via the \nion{Rb}{i}
resonance lines at 7800 and 7947\,\AA\ at the expected low Rb abundances
is possible for cool dwarfs and giants, as our exploratory synthetic
spectra indicated. Stars for observation at high-spectral resolution
were selected from Schuster \& Nissen's (1988)\nocite{SN88} catalog of dwarfs,
and from Pilachowski, Sneden, \& Kraft's (1996)\nocite{PSK96}
list of giants. Emphasis
was placed on metal-poor stars such that the metallicity range
$-2 <$ [Fe/H] $< -0.5$ is well represented but metallicities [Fe/H] $> - 0.5$
are poorly represented. The following sections describe the observations,
the method of analysis, the results, and present an interpretation of the
Rb abundances relative to the  abundances of other elements (Fe, Y, Zr, and Ba)
obtained here.

\section{OBSERVATIONS AND DATA REDUCTION}

The program stars are listed in Table 1.  They
comprise 32 G and K dwarfs and subgiants with
metallicities of $-1.8 <$ [Fe/H] $< 0.0$ and 12 G and K
giants with metallicities of $-2.0 <$ [Fe/H] $< -0.6$.
The observations were made at the McDonald Observatory
with the 2.7-m telescope and 2dcoud\'e echelle spectrometer
(Tull et al. 1995\nocite{TMS95}).  All the program stars were observed at
the F3 focus at a resolving power of R = 60,000 and, in addition,
eight of the brighter stars were also observed at the F1
focus at a resolving power of 200,000.
In order to minimise the influence of cosmic rays, two
observations in succession, rather than one longer
observation, were generally made of each star.
Four different detectors were used for the observations
with resolving power 60,000: a Texas Instruments CCD
with 15\,$\mu$m$^2$ pixels in a 800$\times$800 format, a Tektronix
CCD with 27\,$\mu$m$^2$ pixels in a 512$\times$512 format, the
Goddard Advanced Imaging System CCD with 21\,$\mu$m$^2$
pixels in a 2048$\times$2048 format, and a Tektronix
CCD with 24\,$\mu$m$^2$ pixels in a 2048$\times$2048 format.
The first two of these CCDs provided partial coverage of
the wavelength interval $\sim$\,5500 -- $\sim$\,8000\,\AA\
with large gaps between the end of one spectral order
and the beginning of the next.  The last two CCDs,
which are much larger, provided nearly complete coverage from
$\sim$\,4000 -- $\sim$\,9000\,\AA; coverage was complete from the start
of this interval to 5600\,\AA\ and substantial, but
incomplete, from 5600\,\AA\ to the end of the interval.
The Tektronix CCD with 24\,$\mu$m$^2$ pixels in a
2048$\times$2048 format was used for the 200,000
resolving-power observations, which provided partial
coverage of the region from $\sim$\,5500 -- $\sim$\,8000\,\AA.
The typical signal-to-noise ratio of the extracted one-dimensional
spectra is between 100 and 300 at red and near-infrared
wavelengths for the 60,000 resolving-power observations, while it
is typically between 100 and 250 at the same wavelengths for
the 200,000 resolving-power observations.



%
%


The only accessible Rb lines in stellar spectra are the
two \nion{Rb}{i} resonance lines at 7800.3 and 7947.6\,\AA.  Since
these lines are typically weak we concentrated our
attention on the stronger 7800.3\,\AA\ line and did not
pursue the 7947.6\,\AA\ line, which is half as strong as the
7800.3\,\AA\ line.  Figure~\ref{spec} shows examples of the 7800.3\,\AA\
\nion{Rb}{i} line in the program stars; as may be seen, the line
is partially blended with a stronger \nion{Si}{i} line at 7800.0\,\AA.

The data were processed and wavelength calibrated in a
conventional manner with the IRAF\footnote{
IRAF is distributed by the National Optical Astronomical Observatories,
which is operated by the Association for Universities for Research in
Astronomy, Inc., under contract to the National Science Foundation.}
package of programs
on a SPARC5 workstation.  The 7800.3\,\AA\ \nion{Rb}{i} line
was analyzed by spectrum
synthesis because of the presence of the
\nion{Si}{i} line.  Equivalent widths were measured for lines
of other elements used in the investigation; these
were lines of \nion{Fe}{i} and \nion{Fe}{ii} and the available
lines (\nion{Y}{ii}, \nion{Zr}{i},
\nion{Ba}{ii}, and \nion{Nd}{ii}) of other heavy
elements besides Rb.  Lines suitable for measurement
were chosen for clean profiles, as judged by inspection
of the solar spectrum at high resolution and signal-to-noise
ratio (Kurucz et al. 1984\nocite{KFB84}), that could be reliably
measured in all, or most of, the program stars.
Moore, Minnaert, \& Houtgast (1966)\nocite{MMH66} was our primary source of line
identification.  The equivalent width of each line
was measured with the IRAF measurement option most
suited to the situation of the line; usually this was
the fitting of a single, or multiple, Gaussian
profile to the line profile.  These equivalent widths
were measured from the spectra with 60,000 resolving power;
the 200,000 resolving-power spectra were not used because
their much more limited wavelength coverage excluded
most lines of interest.  Table 2 gives basic information
for the \nion{Rb}{i} line and the lines of the other elements.
The equivalent widths of the lines
are available at JT's World Wide Web site
(http://anchor.as.utexas.edu/tomkin.html).

The spectrum of an asteriod (Iris), observed with the
same instrumental setup as that used for the 60,000
resolving-power observations of the program stars and reduced
and measured in the same manner, provided solar
equivalent widths for these lines.

\section{ANALYSIS}

An LTE model-atmosphere abundance analysis was made relative
to the Sun.  Here we briefly discuss the selection of atomic
data for the lines, the model atmospheres, and the abundance
determinations for Rb and the other heavy elements.

\subsection{Line Data}

The lines used for the abundance determinations are given in Table 2.
Our first choice for $gf$-values for the lines was
modern laboratory $gf$-values; basic data for the lines,
including the $gf$-values and their sources, are given
in Table 2.  In particular we note that the $gf$-value of the
7800.3\,\AA\ \nion{Rb}{i} line,
log\,$gf$ = +0.13$\pm$0.04 (Wiese \& Martin 1980\nocite{WM80}),
is reliably determined.  For some lines, for which reliable
$gf$-values are not available, we used solar $gf$-values instead.
These were calculated with the solar equivalent widths given in
Table 2, the solar atmosphere of
Holweger \& M\"{u}ller (1974)\nocite{HM74}, a microturbulence of 1.15\,\kmsec\
(Tomkin et al. 1997\nocite{TEL97}), and the abundances given in Table 2.
The solar equivalent widths were measured from the spectrum of
the asteriod Iris.

The adoption of the laboratory $gf$-values for the \nion{Fe}{i} lines
follows the prescription of Lambert et al. (1996)\nocite{LHL96}.  In particular,
we make a small correction (see Table 2) to the $gf$-values of
May, Richter, \& Wichelmann (1974)\nocite{MRW74} to normalize them to those of
O'Brian et al. (1991)\nocite{OWL91}
and Bard, Kock, \& Kock (1991)\nocite{BKK91}.  No correction
is needed to put the solar \nion{Fe}{i} line $gf$-values on the same scale
as the laboratory $gf$-values; we find:
log $gf$(solar) -- log $gf$(lab) = +0.03$\pm$0.05 (4 lines,
6 $gf$-values).  Accurate experimental $gf$-values are not
available for our \nion{Fe}{ii} lines, therefore we have used solar
$gf$-values for these lines.

Although the three \nion{Zr}{i} lines all have modern laboratory $gf$-values
(Bi\'emont et al. 1981\nocite{BGH81}), we find their solar $gf$-values
are significantly larger than the laboratory ones:
log $gf$(solar) -- log $gf$(lab) = +0.41$\pm$0.04.
Our adopted solar Zr abundance --- \eps{Zr} = 2.60 (Anders \&
Grevesse 1989\nocite{AG89}) --- is not a factor in this discrepancy
because it is based on Bi\'emont et al.'s $gf$-values and
is very similar to the value --- \eps{Zr} = 2.56 --- Bi\'emont et al.
derived in their own investigation of the solar Zr abundance.
Some of the discrepancy is attributable to the use of different
ionization potentials for \nion{Zr}{i}; we have used the accurate value of
6.634\,eV (Hackett et al. 1986\nocite{HHM86}) while Bi\'emont et al. must have
used the significantly higher old value of 6.84 (Allen 1973\nocite{A73}).  This
accounts for 0.20\,dex of the discrepancy.  Line-to-line variation
of individual line abundances may also account for some of the
discrepancy; the individual line abundances Bi\'emont et al.
derive for these three lines are on average 0.06\,dex larger
than the average Zr abundance they determine from the \nion{Zr}{i} lines.
Although the remaining 0.15\,dex discrepancy is not readily
accounted for, the manner in which these three lines strengthen
in the cooler program stars leaves no doubt that they are
low excitation lines of a neutral species and is thus consistent with their
identification as \nion{Zr}{i} lines.  In order to minimize the influence
of whatever is causing the residual discrepancy between the solar
and laboratory $gf$-values for these lines, we have chosen to adopt
the solar $gf$-values for them in our analysis.

Two \nion{Y}{ii} lines (5200.42 and 5402.78\,\AA) were rejected when it
became evident, during an initial abundance analysis of all
the stars, that they gave significantly larger Y abundances
than the other \nion{Y}{ii} lines in many of the stars.

Atomic data and solar equivalent widths for the lines are
given in Table 2.

\subsection{Model Atmospheres}

Plane parallel, line blanketed, flux constant, LTE MARCS model
atmospheres, which derive from a development of the programs
of Gustafsson et al. (1975)\nocite{GBE75}, were used for the abundance analysis.

The determination of parameters (effective temperature,
surface gravity, metallicity, and microturbulence) for the model
atmospheres was done in two steps.  First we chose preliminary
parameters for each star which were used to calculate an initial
set of model atmospheres.  We then used the initial model atmospheres
and the equivalent widths of the \nion{Fe}{i} and \nion{Fe}{ii} lines
to iteratively adjust the atmospheric parameters to determine an
adopted set of parameters and atmospheres that are consistent with the
\nion{Fe}{i} and \nion{Fe}{ii} line data.  We now briefly describe these
two steps.

\subsubsection{Choice of Preliminary Atmospheric Parameters}

Str\"omgren photometry provided the primary means of determining
preliminary parameters for the dwarfs and subgiants,
which are mostly from the catalogue
of Schuster \& Nissen (1988)\nocite{SN88} supplemented by three stars from
Carney et al. (1994)\nocite{CLL94} and four stars from the Bright Star Catalogue
(Hoffleit \& Jaschek 1982\nocite{HJ82}).
The stars are intrinsically faint and nearby;
an examination of their parallaxes (see below) shows they
are all within 100 pc, except for two stars which are at 111$\pm$17
and 137$\pm$18\,pc.  The interstellar reddening of the
stars is therefore negligible (Schuster \& Nissen 1989a\nocite{SN89a}) so we
have not corrected their indices.

Our chief source of preliminary effective temperatures
for the dwarfs and subgiants was the
color index $b-y$ coupled with the \teff\ vs. $b-y$ calibration of
Alonso, Arribas, \& Mart\'{\i}nez-Roger
(1996a, \nocite{AAM96a} equation 9), who used the infrared flux method to
determine effective temperatures for metal-deficient stars.
For the components of the visual binary HD\,23439, which do not
have their own $b-y$, and for HD\,150281, which also does not have
$b-y$, we took the effective temperatures from
Carney et al. (1994)\nocite{CLL94};
for 61~Cyg~A and B, which are too cool for the applicable range
of Alonso et al.'s calibration, we took the effective temperatures
from Alonso, Arribas, \& Mart\'{\i}nez-Roger (1996b)\nocite{AAM96b}.

Initial surface gravities for the dwarfs and
subgiants were determined by the relations
$g \propto M/R^2$ and $L \propto R^2\teff ^4$
where $M$ is the stellar mass, $R$ the radius, and $L$ the
luminosity, with the luminosities being set by the
Hipparcos parallaxes.  We followed the
prescriptions of Nissen, H{\o}g, \& Schuster (1997)\nocite{NHS97}, who
have successfully applied this method to determine
surface gravities for 54 metal-poor stars.  Most of the stars
have Hipparcos parallaxes and all of these parallaxes are of
sufficient accuracy; the largest uncertainty ($\sigma /\pi$) is
0.15 with most uncertainties being much smaller than this.
For the small number (five)
of stars without Hipparcos parallaxes we took
trigonometric parallaxes from Gliese \& Jahreiss (1979)\nocite{GJ79} or
determined photometric parallaxes from $V$ and $M_V$, with $M_V$
estimated from the $uvby$ photometry and the recipes of Nissen \&
Schuster (1991)\nocite{NS91}.

Preliminary metallicities for the dwarfs and subgiants
were estimated from the $uvby-\beta$ photometry
and the calibration of Schuster \& Nissen (1989b, \nocite{SN89b} equation 3).
For a small number (five) of stars the results of Carney et al.
(1994)\nocite{CLL94} or Alonso et al. (1996b)\nocite{AAM96b} were used instead.

An initial microturbulence of 1.0\,\kmsec, which is representative of
dwarfs in this temperature range (Feltzing \& Gustafsson 1998\nocite{FG98}),
was used for the dwarfs and subgiants.

For the giants, which are taken from
Pilachowski et al.'s (1996)\nocite{PSK96}
medium-resolution spectroscopic investigation of Na abundances in
metal-poor giants, we adopted Pilachowski et al.'s atmospheric
parameters as initial parameters.  Pilachowski et al.'s
effective temperatures and gravities are photometrically based
as modified by their spectroscopic results, while their
metallicities and microturbulences are from their spectroscopic
analysis.

\subsubsection{Determination of Adopted Atmospheric Parameters}

Model atmospheres were computed
using the MARCS code (Gustafsson et al. 1975\nocite{GBE75}).
Those for the dwarfs and subgiants were calculated by interpolation
in a grid of MARCS models, which spanned the range of
dwarf and subgiant parameters and was provided by B. Edvardsson.
The models of Pilachowski et al. (1996)\nocite{PSK96}, who also used MARCS
models for their abundance analysis, were used as preliminary
models for the giants and were provided by C. Sneden; iterations
of the giant models with modified atmospheric parameters were
calculated directly using the MARCS code.

Models with the preliminary parameters and the line analysis code
MOOG (Sneden 1973\nocite{S73}) were then applied to the equivalent width data
for the \nion{Fe}{i} and \nion{Fe}{ii} lines.
Trends of the \nion{Fe}{i} abundances with line excitation potential
were used to check the preliminary effective temperatures and
trends of the \nion{Fe}{i} abundances with equivalent width were used
to check the preliminary microturbulences.  Where necessary
revised parameters were determined, new models calculated,
and a new round of abundance calculations done for the \nion{Fe}{i}
and \nion{Fe}{ii} lines.  Next the Fe abundances from this
round of calculations were used to calculate a new set of
models and do another round of abundance calculations for the
Fe lines. In a final round of calculations the surface gravities
were adjusted, and new models were calculated, so the \nion{Fe}{i}
and \nion{Fe}{ii} lines gave the same Fe abundance.  These final
adopted parameters for the program stars are given in Table 1.

The adopted parameters are generally only moderately different
from the initial parameters.  For the dwarfs and subgiants the
analysis of the Fe lines led to revised effective temperatures
for 13 stars and an average temperature increase for these stars
of 140$\pm$75\,K ($\sigma$ of the individual differences).
The revised gravities of the dwarfs and
subgiants tend to be lower than the preliminary gravities, but
the inconsistency is small; for the 28 dwarfs and subgiants with
measurements of both \nion{Fe}{i} and \nion{Fe}{ii} lines, for which
spectroscopic gravities can thus be determined, the average downward
revision of the log~$g$ is 0.12$\pm$0.12 ($\sigma$ of the 
individual differences).  This suggests that
there is no serious inconsistency between the gravities of these
stars and the four remaining dwarfs and subgiants for which we
adopt preliminary gravities because they have no \nion{Fe}{ii} lines.

It is of interest to see how the differences that we find
between our spectroscopic gravities and the preliminary
Hipparcos-based gravities compare with what Allende Prieto
et al. (1999)\nocite{AGL99} found in a thorough examination,
of spectroscopic gravities, for nearby stars,
taken from the literature versus Hipparcos-based
gravities.  We confine the comparison of
log~$g_{\rm spec}$ -- log~$g_{\rm Hipp}$
for our results and theirs to the temperature range (4900 -- 5500\,K)
of our dwarfs and subgiants.  (This range excludes the
four coolest dwarfs and subgiants because they do not have
any measured \nion{Fe}{ii} lines.)  As mentioned earlier,
we find an average difference
log~$g_{\rm spec}$ -- log~$g_{\rm Hipp}$ = --0.12$\pm$0.12
(28 stars, $\sigma$ of individual differences) for our
dwarfs and subgiants, while Allende Prieto et al. find an
average difference of --0.26$\pm$0.29 (9 stars) for their
sample of stars.  Our results and Allende Prieto et al.'s
thus both show that the spectroscopic gravities tend to
be smaller than the Hipparcos gravities.  Also, the
consistency between the spectroscopic and Hipparcos gravities
of our stars is somewhat better than it is for Allende
Prieto et al.'s sample of stars.\footnote{
We have not determined Hipparcos-based gravities
for the giants because, with the exception of Arcturus, they
are much more remote than the dwarfs and subgiants.  Although
they all have Hipparcos parallaxes, the errors in the
parallaxes are comparable to the parallaxes for most of them.}

Revisions of effective temperature were required for five of
the giants; the average temperature increase for these
stars was 100$\pm$140\,K.  All the giants have measurements
of both \nion{Fe}{i} and \nion{Fe}{ii} lines thus allowing revision of their
preliminary gravities for all 12 stars; the average change
of log~$g$ was -0.04$\pm$0.37.  The adopted effective temperatures
and gravities of the giants thus are in good agreement with
the preliminary values (Pilachowski et al. 1996\nocite{PSK96}).
Our results, which use higher resolution spectra
and more numerous Fe lines than those of Pilachowski
et al., thus confirm their results.  Our Fe abundances
are also in good agreement; the
average difference between our [Fe/H] determinations and
theirs is -0.07$\pm$0.10.

We conclude this section with a brief discussion of the potential
influence of non-LTE on the effective temperatures derived from the
\nion{Fe}{i} line excitation.  As remarked earlier, the primary source
of our initial effective temperatures (for the dwarfs and subgiants)
was $b-y$ and the \teff\ vs. $b-y$ calibration of
Alonso, Arribas, \& Mart\'{\i}nez-Roger
(1996a, \nocite{AAM96a}), where the effective temperatures in their
calibration were determined by the infrared flux method.
The $b-y$ based effective temperatures are thus free of non-LTE effects.
One way to estimate the possible influence
of non-LTE on the effective temperatures derived from the
excitation of \nion{Fe}{i} lines, therefore, is to consider the difference
between the excitation-based temperatures and the $b-y$-based
temperatures.  Of course non-LTE effects are not the only
possible source of such a difference so this check is indicative,
rather than conclusive.  The average difference
\teff\ (\nion{Fe}{i}) $-$ \teff\ ($b-y$) = +45$\pm$68\,K
($\sigma$ of the individual differences, 20 stars), where the
calculation includes not only stars whose initial effective
temperatures were revised, but also stars for which no revision
was necessary - as long as they had enough \nion{Fe}{i} lines
to define the line excitation.  This difference is small and
indicates that any non-LTE influence on the determination of
temperatures from the \nion{Fe}{i} lines is minor.

\subsection{Abundance Determinations for Rb and the Other Heavy Elements}

Abundances were determined by matching the observed line
strengths and theoretical line strengths calculated by MOOG
with the adopted model atmospheres.  As remarked earlier,
the \nion{Rb}{i} line was treated by means of spectrum synthesis, while
the lines (\nion{Y}{ii}, \nion{Zr}{i}, \nion{Ba}{ii}, and \nion{Nd}{ii})
of the other heavy elements were treated by means of equivalent widths.

The spectrum synthesis of the 7800.29\,\AA\ \nion{Rb}{i} line includes the
hyperfine structure of the $^{85}$Rb and $^{87}$Rb isotopes, each of
which is split into two components, and the blending
\nion{Si}{i} line at 7800.00\,\AA.  The accurately known
wavelengths and relative line strengths of the hyperfine
structure components were taken from Lambert \& Luck's (1976)\nocite{LL76}
analysis of Rb in Arcturus.  We adopted a terrestrial
abundance ratio ($^{85}$Rb/$^{87}$Rb = 3) for the Rb isotopes.
Although in principle it would be desirable to make direct measurements
of the stellar isotopic Rb abundances from the exact shape
of the RbI line profile, in practice extreme departures from the
terrestrial isotope ratio are required before there is appreciable
distortion of the line profile.  Lambert \& Luck, for example,
found that for their spectra, which had a resolving power of
195,000 that is similar to the high resolution spectra of the
present investigation, the $^{85}$Rb/$^{87}$Rb ratio had to be as low as
1 or as high as 10 to cause even a small variation of the line
profile; they concluded that the isotope ratio in Arcturus is
terrestrial with a large uncertainty.  Direct determination of Rb
isotopic abundances, therefore, is not the thrust of our
investigation, although we note that comparison of the observed
and synthesised spectra of the Rb line in our program stars
does not show any variations attributable to non-terrestrial isotope
ratios.  We also note that the indeterminacy of the Rb isotope ratios
does not interfere with the measurement of the elemental
Rb abundances; the \nion{Rb}{i} line is weak in all the program stars
so the Rb abundances it provides are not affected by
changes of the isotopic mixture.

The synthesis of the \nion{Rb}{i} line also includes the \nion{Si}{i} line
to the blue.  No reliable experimental oscillator strength is
available for the \nion{Si}{i} line so a solar oscillator strength
(log~$gf$ = -0.65) was adopted.  The instrumental and
macroturbulent broadening, as well as thermal and
microturbulent broadening, were included in the synthesis.
The macroturbulent broadening was set by matching the profile
of the clean nearby \nion{Ni}{i} line at 7797.6\,\AA.

Spectrum synthesis of the solar \nion{Rb}{i} line, using the solar
model of Holweger \& M\"uller (1974)\nocite{HM74} and the Kurucz et al.
(1984)\nocite{KFB84} solar atlas, provides a solar Rb abundance
\eps{Rb} = 2.60$\pm$0.07, which is the same as the
Rb abundance in Anders \& Grevesse's (1989)\nocite{AG89} compilation
of solar abundances.  We note, however, the discrepancy
between the photosperic abundance and the somewhat lower
meteoritic abundance of 2.40$\pm$0.03
(Anders \& Grevesse 1989\nocite{AG89}).  Although it might be speculated
that the relatively low melting and boiling points of
Rb (39 and 688\,C, respectively) may make it behave like a
volatile element and so explain the low meteoritic abundance,
this does not appear to be the case.  Potassium, which is
isoelectronic with Rb and has only slightly higher melting
and boiling points (63 and 759\,C, respectively), shows no
discrepancy of its photospheric and meteoritic abundances,
which are \eps{K} = 5.12$\pm$0.13 and 5.13$\pm$0.03,
respectively.  Although the discrepancy is a potential
source of concern, we note that there is no evidence of the
\nion{Rb}{i} line being affected by an unknown blend; in particular,
in our program stars the strengthening of the line with
decreasing effective temperature is consistent with the
behavior of a resonance line of a heavy-element neutral
species.

The observed and synthesised \nion{Rb}{i} line profiles for a sample
of stars are shown in Figure~\ref{spec}.  Table 3 gives the abundances of
Rb and the other heavy elements.

The Rb abundances derived from observations made at the F1
focus (resolving power = 200,000) and the F3 focus
(resolving power = 60,000) are highly consistent;
for stars observed at both foci the average difference
between the F1- and F3-based [Rb/H] is --0.03$\pm$0.01 (s.e., five stars).
For bright stars, such as Arcturus and $\mu$ Cas, the greater
detail provided by the higher resolution F1 observations allows
for more precise determination of the Rb abundance.  In fainter
stars there is not much to choose between the F1 and F3 observations
because the greater spectral detail of the F1 observations tends
to be counterbalanced by their lower signal-to-noise ratio.

The two main sources of errors in the abundances are measurement
error and analysis error caused by errors in the adopted model
atmosphere parameters.  The scatter of the abundances provided
by individual lines of the same species, which are caused by
measurement errors of the equivalent widths and, to a lesser
extent, by errors in the line oscillator strengths, is a good
guide to measurement error.  This scatter, as measured by the
standard deviation of the individual line abundances, is given
in Table 3.  (Although the standard deviations of the abundances
from individual lines are larger than the standard deviations of
the mean abundances, the heavy element abundances are based on
only a few lines for each element --- see Table 2 --- so we prefer
to consider the standard deviations of the individual line
abundances.)  For Rb, whose abundance is based on spectrum
synthesis of the 7800\,\AA\ \nion{Rb}{i} line, Table 3 gives errors estimated
from the fit of the observed and synthesised spectra.  Inspection of
Table 3 shows the measurement-related abundance errors range
up to $\pm$0.15\,dex with larger errors in a few cases; a representative
error in the [X/H] abundances is $\sim \pm$0.07 dex, while a
representative error of the [X/Fe] abundances is $\sim \pm$0.1 dex.

Estimated errors in the adopted effective temperatures are
between $\pm$50\,K, for stars with a good selection of \nion{Fe}{i} lines
providing a well-determined excitation temperature, and
$\pm$100\,K, for stars for which we adopted color-based effective
temperatures.  Representative errors in the adopted log~$g$
and metallicities are $\pm$0.2 and $\pm$0.1\,dex, respectively.
A representative uncertainty in the microturbulence is
$\pm$0.5\,\kmsec, although we note that the Rb abundances provided
by the weak \nion{Rb}{i} line have little, or no, microturbulence
dependence and that because of the metal deficiency of most of
the program stars the abundances of most other elements also
have only a small, or negligible, microturbulence dependence.
Adopting a representative effective temperature error of
$\pm$100\,K and the stated errors of the other parameters,
we find that for a typical dwarf the combined effects of these
errors change the Fe abundance (from \nion{Fe}{i} lines) by $\pm$0.11\,dex.
The corresponding figure for a typical giant is $\pm$0.17\,dex.
For the heavy elements the abundance of the element relative
to Fe, [El/Fe], holds the most interest.  This ratio is less
dependent on the atmospheric parameters than the absolute
abundance.  In the typical dwarf the combined effects of the
errors in the atmospheric parameters change [El/Fe] by
$\pm$0.04 (Rb), $\pm$0.05 (Y), $\pm$0.05 (Zr), $\pm$0.11 (Ba),
and $\pm$0.06\,dex (Nd).  The corresponding figures for a typical
giant are: $\pm$0.07 (Rb), $\pm$0.06 (Y), $\pm$0.00 (Zr),
$\pm$0.14 (Ba), and $\pm$0.07\,dex (Nd).  We estimate representative
total errors, caused by measurement error and errors in the model
atmosphere parameters together, to be 0.1 -- 0.2 dex in [Fe/H]
and 0.1 -- 0.2 dex for [El/Fe].

We now briefly consider how the Fe and Rb abundances of the
dwarfs and subgiants would change if the preliminary effective
temperatures, which are mostly based on the infrared flux
method (Alonso et al. 1996a\nocite{AAM96a}), and preliminary
gravities, which are mostly Hipparcos-based, were used instead
of the adopted effective temperatures and gravities.
As discussed earlier, the preliminary effective temperatures
were revised upward by an average of 140$\pm$75\,K
($\sigma$ of the individual differences) for 13 of the
dwarfs and subgiants, while no revisions of the preliminary
effective temperatures were made for the other 19 dwarfs
and subgiants.  Use of the lower preliminary effective
temperatures for these 13 stars would decrease their Fe
abundances (from \nion{Fe}{i} lines) by an average of 0.14\,dex,
while their [Rb/Fe], as set by \nion{Rb}{i} and \nion{Fe}{i}
lines, would increase by an average of 0.04\,dex.
The adopted gravities for 28 of the dwarfs and subgiants
are spectroscopically determined, while those of the other
four dwarfs and subgiants, for which we could not determine
spectroscopic gravities, are the preliminary
gravities.  Use of the preliminary gravities, instead of the
adopted spectroscopic gravities, for these 28 stars would not
change their Fe or Rb abundances significantly; the adoption
of the preliminary log $g$, which are 0.12$\pm$0.12\,dex
($\sigma$ of the individual differences) higher on
average than the spectroscopic log $g$, in place of the
spectroscopic log $g$ would change the [Fe/H] and [Rb/Fe]
from neutral lines by +0.01 and 0.00\,dex, respectively,
on average.

\section{RESULTS}

\subsection{Comparison with the Literature}

Before we consider the Rb and other heavy-element abundances,
we compare our results with those published in the literature.
First we consider the results for [Fe/H].  Figure~\ref{Fecomp} compares
the [Fe/H] determinations for stars in common to this study
and earlier high signal-to-noise ratio, high resolution studies.
The comparison is not exhaustive, but does include all recent
studies (since 1990) which have two, or more, stars in common
with the present study.  The agreement of the [Fe/H] determinations
is good over most of the metallicity range and, although the
[Fe/H] of this study tend to be slightly more negative than
the literature [Fe/H] in the most metal-deficient stars, the
overall agreement is not unsatisfactory.

Previous studies which, to our knowledge, have determined Rb
abundances for stars in common with those of the present
investigation are M\"ackle et al.'s (1975a)
\nocite{MHG75a} study of Arcturus
and Gratton \& Sneden's (1994)\nocite{GS94} study of heavy-element abundances
in metal-poor stars.  In Table 4 we compare our [Rb/Fe] with
those of the two earlier investigations.  Because our Rb
abundances are based on the 7800\,\AA\ \nion{Rb}{i} line, while M\"ackle
et al.'s (1975a)\nocite{MHG75a}
abundance is based on both the 7800 and 7947\,\AA\
lines, we also include in the Table their abundance for the
7800\,\AA\ \nion{Rb}{i} line alone (M\"ackle et al. 1975b)\nocite{MGG75b}.
(Gratton \& Sneden's results are based only on the 7800\,\AA\ line.)
In order to make the Rb abundances of our and the earlier
studies directly comparable we have adjusted the Rb
abundances of the earlier studies to reflect the values they
would have if the stellar parameters (\teff, log~$g$, [M/H], and
$\xi$) used in the earlier studies had been the same as those
used here.  For Arcturus the difference between our [Rb/Fe]
and M\"ackle et al.'s is only --0.03 dex --- pleasingly small and
not unexpected for the case of such a bright star.  For the
three stars that we have in common with Gratton \& Sneden
we note that both investigations determined
Rb abundances for two of the stars (HD\,64606 and 187111), but were
only able to determine upper limits for the third star
(HD\,122956).  The differences between our [Rb/Fe] and
Gratton \& Sneden's are --0.25 (HD\,64606), --0.09 (HD\,122956),
and +0.17 (HD\,187111).  That these differences are much larger
than in the case of Arcturus can be ascribed to the fact
that all three stars are quite metal-deficient and have only
a weak or undetectable Rb line.  We estimate that for these
three stars the uncertainty in the Rb abundance associated
with fitting our observed and synthetic spectra of the \nion{Rb}{i}
line is 0.1\,dex for HD\,64606 and 187111 and 0.2\,dex for
HD\,122956; Gratton \& Sneden's Rb abundances must be subject
to similar uncertainties also.  We conclude, therefore,
that our [Rb/Fe] and Gratton \& Sneden's are probably the
same to within the errors of measurement for these three stars.

\subsection{The Abundances of Rb, Y, Zr, Ba, and Nd}

As is customary, the abundances in Table 3 are plotted as [el/Fe] against
[Fe/H], in Fig.~\ref{Rb.Y}, \ref{Zr.Ba}, and \ref{Nd}
to reveal trends in the relative
abundance of element el and iron.
Two points are immediately apparent: (i) three stars are unusually
rich in the heavy elements --- note especially the [Y/Fe] ratios of
HD\,23439A and B, and BD\,+5$^{\rm o}$\,3640 which we shall dub CH stars,
and (ii) the distinctive behavior of [Rb/Fe] in the metal-poor
stars --- [Rb/Fe] $>0$ when the other heavy elements show [el/Fe] $\simeq 0$.
Before commenting on these striking results, we compare our results
for the Y, Zr, Ba, and Nd abundances with results in the literature.

Previous extensive abundance determinations of heavy (and other) elements in
metal-poor stars have shown that the run of [el/Fe] against [Fe/H] is
smooth down to about [Fe/H] = --2 with  a `cosmic' scatter
less than the scatter that results from the errors of measurement. Cosmic
scatter is present for more metal-poor stars but our sample is devoid of
such stars. Therefore, our results are expected to agree well with previous
studies despite the lack of a complete overlap in stellar samples. Key
papers reporting results on heavy elements are Zhao \& Magain
(1991)\nocite{ZM91} and Gratton \& Sneden (1994)\nocite{GS94}
with reviews by Wheeler, Sneden, \& Truran (1989)\nocite{WST89},
Lambert (1989)\nocite{L89}, and McWilliam (1997)\nocite{M97} amongst others.
Our results for [Y/Fe], [Ba/Fe], and [Nd/Fe] are in excellent agreement
with previous results, for example, Zhao \& Magain (1991)\nocite{ZM91} and
Gratton \& Sneden (1994)\nocite{GS94} find [Y/Fe] $\simeq -0.1 $
at [Fe/H] = --1 with the relative underabundance of Y increasing to about
0.25 at [Fe/H] = --2 which agree well with Figure~\ref{Rb.Y}. A discrepancy
appears when comparing results for [Zr/Fe]. Zhao \& Magain (1991)\nocite{ZM91}
and Gratton \& Sneden (1994)\nocite{GS94} report [Zr/Fe] $\simeq +0.2$ for
[Fe/H] in the range of --1 to --2 but our results
(Figure~\ref{Zr.Ba}) show [Zr/Fe] to be
consistently less than zero: a difference in [Zr/Fe] of about 0.3 to 0.4\,dex
relative to the previous studies. This difference is most probably due to
our exclusive use of \nion{Zr}{i} lines.
Brown, Tomkin, \& Lambert (1983)\nocite{BTL83}
found that \nion{Zr}{i} lines in mildly metal-poor giants gave a clear Zr
underabundance which was plausibly attributed to non-LTE effects such as
over-ionization of Zr atoms to Zr$^+$ ions. For metal-poor stars, Gratton \&
Sneden remark that their selection of \nion{Zr}{i} lines gives a systematically
lower Zr abundance than the \nion{Zr}{ii} lines: the difference of
--0.16$\pm$0.05\,dex would account in part for our largely
negative values of [Zr/Fe].

Our results clearly show a relative overabundance of Rb in metal-poor
stars: the mean value [Rb/Fe] = +0.23$\pm$0.02 (s.e.) is found from nine stars
with [Fe/H] $< -1$, excluding the three CH stars.
%
%
This is consistent with the four measurements
reported by Gratton \& Sneden (1994)\nocite{GS94} and with their
six upper limits to [Rb/Fe].

Non-LTE effects such as overionization warrant consideration.
As a guide to the non-LTE effects on Rb,
we consider those calculated for lithium, another alkali. Carlsson et al.
(1994)\nocite{CRB94} predict that the LTE abundances from the \nion{Li}{i}
6707\,\AA\ resonance doublet  require correction by not more than
0.03\,dex for non-LTE effects, a negligible correction in the present
and almost all contexts. The abundances of Li and Rb are
similar: lithium has the abundance \eps{Li} $\simeq 2.2$ in
the warmer dwarf stars comprising the Spite plateau and Rb declines from
\eps{Rb} = 2.6 at solar metallicity to 1.9 at [Fe/H] = --1
and to 0.8 at [Fe/H] = --2. The key point is that optical depth effects in
lines and continua are slight for both elements. Rb is probably
more affected by photoionization because the atom's ionization potential
is 4.18\,eV versus 5.39\,eV for the lithium atom. Photoionization of Rb
will be enhanced relative to the rate for Li but collisional ionization
rates will also be enhanced. The different wavelengths of the resonance
(and excited) atomic lines for Rb and Li will introduce no more than
slight differences in the non-LTE corrections. If non-LTE effects
were large for Rb, we would anticipate that dwarfs and giants of the
same metallicity would yield systematically different Rb abundances.
This is not the case: four dwarfs with [Fe/H] in the range --1 to --2 give
a mean [Rb/Fe] = +0.28$\pm$0.04 (s.e.) and five giants in the same [Fe/H]
range give the mean [Rb/Fe] = +0.18$\pm$0.02 (s.e.).
(The mean for the dwarfs excludes the three CH stars.)  Although a non-LTE
analysis for Rb would be of interest, we suggest that our results derived
from LTE analyses are not substantially different from  non-LTE
results.

Scatter of the [el/Fe] results at a given [Fe/H] is not significantly different
from that expected from the measurement errors.
Obviously, the three stars
over-abundant in the heavy elements and dubbed CH stars are  set
aside  as special cases. The scatter for [Rb/Fe] between [Fe/H]
of --0.5 and --1.0 is small and consistent with the measurement errors. There
is an apparent moderate increase in scatter of [Rb/Fe] below [Fe/H] = --1 but
this is probably again due to the measurement errors because the \nion{Rb}{i}
line is very weak in these metal-poor stars. The results are roughly
consistent with a constant [Rb/Fe] in stars with [Fe/H] $< -1$.

\subsection{The New CH Stars HD\,23439A and B and BD\,+5$^{\rm o}$\,3640}

These three stars, which are all dwarfs, are consistently
overabundant in all of the five heavy elements investigated in
this study.  Figure~\ref{CH} shows a \nion{Zr}{i} line and a \nion{V}{i} line in
HD\,23439A and B and HD\,103095, a non-CH star with otherwise
similar properties.  The much greater strength of the \nion{Zr}{i}
line relative to the \nion{V}{i} line in HD\,23439A and B as
compared with HD\,103095 is evident.
As may be seen in Fig.~\ref{Rb.Y}, \ref{Zr.Ba}, and \ref{Nd}
the heavy-element abundance
enhancements are similar in the three stars.  The average
enhancements for the three stars are: [Rb/Fe] = +0.41,
[Y/Fe] = +0.34, [Zr/Fe] = +0.51, [Ba/Fe] = +0.27, and
[Nd/Fe] = +0.32 (HD\,23439A and BD\,+5$^{\rm o}$\,3640); actual values
of [X/Fe] for the individual stars are given in Table 5.
HD\,23439 is a nearby visual binary composed of a K1V primary
and a K2V secondary; the Hipparcos Catalogue
gives 7.$''$307 and 40.83\,mas for the separation of its components
and parallax, respectively.  These numbers set a lower limit
on the A--B linear separation of 179\,AU.
To the best of our knowledge HD\,23439
is the first case of a binary in which both components have
been found to be CH stars.

We note that HD\,23439B is a single-lined spectroscopic binary
with a period of 48.7\,d and a mass function of 0.0022
(Latham et al. 1988\nocite{LMC88}).  Could the unseen companion
of HD\,23439B be the white dwarf descendant of the
AGB star responsible for the mass
transfer that changed HD\,23439A and B into CH stars?  Perhaps,
but in this scheme it is hard to explain the very similar
heavy-element enhancements of HD\,23439A and B (see Table 5).
Just as today the unseen companion is much closer to component B
than to component A so in the past the putative AGB predecessor
of the unseen companion must also have been much closer to B
than A.  How did the AGB star manage to give the A and B
components the same heavy-element enhancements?  This difficulty
with the AGB-star scenario suggests that the heavy-element
enhancements must be primordial.

\section{RUBIDIUM AND STELLAR NUCLEOSYNTHESIS}

Heavy elements are synthesised by the neutron capture $s$- and $r$-processes.
(In considering elemental abundances, the small contribution from
$p$-processes may be neglected.) Detailed dissection of the isotopic
abundances measured for carbonaceous chondrites has provided 
an isotope by isotope resolution of the abundances into $s$- and $r$-process
contributions (cf. K\"{a}ppeler, Beer, \& Wisshak 1989\nocite{KBW89}).
As is well known,
Ba and Eu are primarily $s$- and $r$-process products respectively:
Cowan (1998)\nocite{C98} estimates that Ba is 85\% an $s$-process
product, and Eu is 97\% a $r$-process product
--- see Gratton \& Sneden (1994)\nocite{GS94}
for similar estimates. Rubidium is of mixed parentage; Cowan gives the
$s$- and $r$-process fractions as 50\% each --- Gratton \& Sneden
provide quite similar estimates (48\% for $s$- and 52\% for
the $r$-process). As noted in the Introduction, the $s$-process contribution
may be broken into a `weak' and a `main' component.  Gratton \& Sneden
divide the 48\% $s$-process Rb contribution into 5\% from the weak and
43\% from the main $s$-process. Our goal is to use the Ba and Eu abundances
as monitors of the $s$- and $r$-processes respectively to predict the
Rb abundances, and then to comment on the consistency between the
predicted and observed Rb abundances.

The other heavy elements considered here are also a mix of $s$- and 
$r$-processes: Cowan gives the following $(s,r)$ \%:  (72, 28) for Y,
(81, 19) for Zr, and (47, 53) for Nd.
Gratton \& Sneden (1994)\nocite{GS94} put the
weak contribution to the total $s$-process as 16\% for Y, 10\% for Zr,
and less than 1\% for Ba, Nd, and Eu. Especially interesting is the roughly
50-50 split for Nd that  matches the split for Rb. Then, the simplest
possible scenario of unvarying yields of $s$ and $r$-process products
over the life of the Galaxy would predict that [Rb/Fe] and [Nd/Fe] would
vary identically with [Fe/H].
Inspection of Fig.~\ref{Rb.Y} and \ref{Nd} shows that
this is not the case.

There are several factors pertinent to the understanding of the run of
Rb and other heavy elements with [Fe/H].
\begin{itemize}
\item
It is now well known that, the distribution of heavy elements
at low [Fe/H] resembles a $r$-process pattern and is not the mix of $s$-
and $r$-processes that prevails at solar metallicities
(Truran 1981\nocite{T81}; Sneden \& Parthasarathy 1983\nocite{SP83};
Wheeler et al. 1989\nocite{WST89}; Lambert 1989\nocite{L89}). There is
evidence that the abundance distribution of the $r$-process was largely
invariant from low metallicities to the present [Fe/H] $\simeq 0$. 
Then, it should suffice in modelling the [el/Fe] vs [Fe/H] relations
to adopt the relative $r$-process abundances that are obtained from
dissection of the measurements on carbonaceous chondrites.
\item
Europium is assigned to the $r$-process: Cowan's (1998)\nocite{C98} resolution
of the meteoritic abundances is 97\% $r$-process and a mere 3\% $s$-process.
With declining metallicity the $s$/$r$ ratio declines. Therefore, 
we may assume that Eu is a $r$-process product throughout the evolution of
the Galaxy.  
The run of [Eu/Fe] against [Fe/H] is taken from McWilliam's (1997)\nocite{M97}
review: [Eu/Fe] = 0 at [Fe/H] = 0 with a smooth transition to [Eu/Fe] $\simeq
0.3$ at [Fe/H] = --1.0 and to the metallicity limit [Fe/H] = --2.5 of interest
to us.  
\item
Red giants enriched in $s$-process heavy elements are likely the
major donors of these elements to the Galaxy's interstellar medium.
Analyses of such red giants of differing [Fe/H] show that the
pattern of $s$-process products has evolved with [Fe/H].
Smith (1997)\nocite{S97}, who
has collated published results, defines Y and Zr as `light' $s$-process
elements (here, ls) and Ba, La, and Ce as `heavy' $s$-process elements
(here, hs). He finds that [hs/ls], which by definition is 0 at [Fe/H] = 0,
increases to [hs/ls] $\simeq$ 0.6 at [Fe/H] = --1.5. 
This evolution of [hs/ls] is attributed to an increase in the average
exposure to neutrons in the He-burning shell of the AGB stars that
are the site of the main $s$-process, as expected on theoretical
grounds.
\item
There is limited evidence also from abundance analyses of red giants
that the Rb abundance relative to other ls elements increases with
decreasing [Fe/H]; Smith's (1997)\nocite{S97} collection of results implies
[Rb/Zr] increases by about 0.7 dex from [Fe/H] = 0 to --1.5. This increase
is attributed to a higher mean neutron density in the He-burning shell
of the metal-poor AGB stars.  As noted by Smith, this increase is expected on
theoretical grounds.
\item
If the magnitude of the
weak $s$-process contributions to the abundances in carbonaceous
chondrites were representative of the $s$-process at all relevant [Fe/H],
the weak $s$-process could be safely dropped from our search for an
explanation of the run of [el/Fe]. A thorough direct check on the
weak $s$-process is not possible because the majority of the
elements between Rb and the
Fe-group are inaccessible spectroscopically. Zinc, which
offers at least a hint of
the behavior of the weak $s$-process,  is assigned 34\% to the $s$-process
and 66\% to the $r$-process in Cowan's breakdown of solar system abundances.
The $s$-process component is essentially entirely due to the weak $s$-process.
This breakdown neglects a possible contribution to Zn from the
sources that contribute the Fe-group elements. In the metallicity range
of interest, [Zn/Fe] = 0.0$\pm$0.15 (Sneden \& Crocker 1988\nocite{SC88}). 
This result, which is barely compatible with the increase of [Eu/Fe]
with declining metallicity, implies a drop in [$s$/Fe] with metallicity
and justifies our neglect of the weak $s$-process contribution to Rb
and other elements.
\end{itemize}

Guided  by these facts, it is possible to predict relative abundances
of the heavy elements including Rb. We begin by considering Ba, Nd, and Eu. 
Eu defines the evolution of the $r$-process products. The observed run of
[Ba/Fe] against [Fe/H] provides the evolution of the hs component of
the $s$-process after a small correction for this element's $r$-process
component based on the Eu abundances and the meteoritic $r$-process
Ba/Eu ratio. The adopted runs of  [Eu/Fe] and [Ba/Fe] against [Fe/H]
are shown in Fig.~\ref{calc.abun}a. Then, it is a simple matter to predict the
run of [Nd/Fe] using the meteoritic 50--50 split into $s$- and $r$-process
contributions. This prediction which is shown too in Fig.~\ref{calc.abun}a is
slightly inconsistent with the observations (Fig.~\ref{Nd}) that
show [Nd/Fe]$\simeq$
0 at all metallicities rather than the predicted [Nd/Fe] = 0.14 at
[Fe/H] $<$ --1. Earlier, we noted that our Nd abundances are
quite consistent with previously published results. This small inconsistency
appears not to have been noted previously.  It is likely that when
the measurement
errors are included the prediction and observations will overlap. 
Note that the prediction uses
the Eu and Ba abundances as well as the meteoritic $s$ to $r$ ratios
for Ba, Nd, and Eu. A change in the Nd ratio from 50\% $s$ and 50\% $r$  to
75\% $s$ and 25\% $r$ reduces [Nd/Fe] to 0.0 for metal-poor stars. 

Yttrium abundances may be predicted from the [Ba/Fe] observations and
Smith's estimates of [hs/ls] from heavy element enriched red giants
such as S and Barium stars.
This prediction is shown in Fig.~\ref{calc.abun}b. The
relative underabundance of Y (i.e., [Y/Fe] $< 0$) results largely
from the steep increase in [hs/ls] with decreasing metallicity
that offsets the increase in the $r$-process contribution. This particular
prediction assumes a meteoritic ratio of 72\% $s$-process and 28\%
$r$-process and makes no attempt to separate main from weak $s$-process
contributions. This prediction matches the observations quite well
(see Fig.~\ref{Rb.Y}). 

The Rb prediction corresponding to the Y prediction which is also shown in 
Figure~\ref{calc.abun}b  does not correspond
to the observed [Rb/Fe] ratios in metal-poor
stars. The limited evidence gathered by Smith suggests that Rb
in heavy element enriched red giants is progressively
overabundant (relative to Zr) in metal-poor giants. This increase is
attributable to a higher neutron density at the $s$-process site in red giants.
If a smooth curve is drawn through Smith's assembled data, the
resulting run of [Rb/Fe] is shown in Fig.~\ref{calc.abun}b. This prediction
is in good agreement with the observations.

\section{CONCLUDING REMARKS}

The principal novel result of our survey of Rb abundances in stars
is that rubidium relative to iron is systematically overabundant
in metal-poor stars.  This increase reflects partly the growth
of the $r$-process abundances relative to iron in metal-poor stars;
we model this increase using observed Eu abundances and the
assumption that the pattern of $r$-process abundances is
solar-like at all metallicities.  A second and major factor accounting
for the increase in the Rb to Fe ratio in metal-poor stars is that
the $s$-process contribution to Rb increases with decreasing
metallicity.  We model this increase using published abundances of Rb
and other heavy elements collated by Smith (1997)\nocite{S97}
for $s$-process enriched red giants that are presumed to be
representative of the donors of $s$-processed material to the
interstellar medium and so to control the chemical evolution of the
Galaxy as far as the $s$-process is concerned.  Two factors
influence the Rb abundance: (i) the total exposure to neutrons at the
$s$-process site, the He-burning shell of an AGB star, increases with
decreasing metallicity of the red giant, and (ii) the neutron
density at the $s$-process site increases with decreasing metallicity.
That (i) is true follows from the observed increase of the relative
abundance ratio of heavy to light $s$-process elements in the
$s$-process enriched red giants.  It is this effect that accounts,
for example, for the drop in [Y/Fe] in metal-poor stars.  That (ii)
is true follows from the limited data on Rb abundances in $s$-process
enriched red giants.  As explained above, neutron density influences
the Rb abundance through the branch in the $s$-process path at
$^{85}$Kr.  In this thoroughly empirical way we account for
the relative enrichment of Rb in metal-poor stars.  In short,
the observed [Rb/Fe] ratios of metal-poor stars are consistent
with the expectation that AGB stars control the input of
main $s$-process products to the Galaxy's interstellar medium.

A serendipitous discovery is the finding that both members of
a visual binary are mild CH stars, HD\,23439A and B, with
$s$-process overabundances relative to other stars of the same
metallicity.  Mass transfer across a binary system is now
thought to account for CH stars and the Barium stars, the higher
metallicity counterparts of the CH stars.  HD\,23439B is a
single-lined spectroscopic binary and the visible star might
have been transformed to a CH star by mass transfer from the
companion, then an AGB star and now a white dwarf.
HD\,23439A appears to be a single star that cannot have captured
significant amounts of mass from a very distant AGB star
orbiting HD\,23439B.  We suggested, therefore, that these CH
stars testify that the halo's interstellar medium
was not entirely chemically homogeneous.  This is not too
surprising given that $s$-process products are injected into the
interstellar medium at low velocity by red giants whereas iron
and other elements are injected at very high velocity by
supernovae.  If the timescale for star formation is shorter than
the timescale for thorough mixing of supernovae and red giant
ejecta, abundance anomalies will result.

\acknowledgements

We thank Bengt Edwardsson for providing the grid of dwarf
and subgiant MARCS model atmospheres, Chris Sneden for
providing Pilachowski et al.'s (1996)\nocite{PSK96} giant MARCS model
atmospheres, and M. Busso and Verne Smith for helpful
discussions.  We also thank Pilachowski et al. for a list
of the stars on their observing program given to us in
advance of publication.
This research has made use of the Simbad database,
operated at CDS, Strasbourg, France.
This work has been supported in
part by NSF grant AST 9618414 and the Robert A. Welch Foundation
of Houston, Texas.

\clearpage
\begin{center}
{\bf Figure Captions}
\end{center}
 
\figcaption{ The \nion{Rb}{i} line at 7800.27\,\AA\
and the adjacent \nion{Si}{i} line
at 7800.00\,\AA\ in three of the program stars.  The top panel
shows a 200,000 resolving-power observation of $\mu$ Cas (G5Vb) and the
middle and bottom panels show 60,000 resolving-power observations
of HD\,65583 (G8V) and HD\,108564 (K2V), respectively.  Synthesised
spectra for [Rb/Fe] of 0.0, 0.3, and 0.6 are shown.
In $\mu$ Cas and HD\,65583 the high-excitation \nion{Si}{i} line
($\chi_{\rm lower}$ = 6.18\,eV) is much stronger than the low-excitation
\nion{Rb}{i} line ($\chi_{\rm lower}$ = 0.00\,eV), but in HD\,108564, which is
$\sim$700\,K cooler than $\mu$ Cas and HD\,65583, the \nion{Rb}{i} line is
slightly stronger than the \nion{Si}{i} line.
\label{spec}}

\figcaption{Comparison of derived [Fe/H] for stars in common to
the present study and earlier high signal-to-noise ratio, high
resolution studies.  Literature shorthand citations are: 
P96 = Pilachowski et al. (1996); S91 =
Sneden, Gratton, \& Crocker (1991);
F98 = Fuhrmann (1998); F97 = Flynn \&
Morell (1997); D93 = Drake \& Smith (1993);
T92 = Tomkin et al. (1992);
P93 = Pilachowski, Sneden, \& Booth (1993); G94 = Gratton \&
Sneden (1994); K97 = King (1997);
G99 = Gratton et al. (1999).
The dashed line shows the locus of identical results.
\label{Fecomp}}

\figcaption{ Plots of [Rb/Fe] and [Y/Fe] versus [Fe/H] for the program
stars.  The three CH stars (HD\,23439A and B and BD\,+5$^{\rm o}$\,3640) are
shown with filled-in squares.  The error bars for [Rb/Fe] are a
quadratic sum of the Rb abundance errors, estimated from the fit
of the observed and synthetic spectra of the 7800\,\AA\ \nion{Rb}{i} line,
and the standard deviations of the Fe abundances from individual lines.
The error bars for [Y/Fe] are a quadratic sum of the
standard deviations of the Y and Fe abundances from individual lines.
\label{Rb.Y}}

\figcaption{ Plots of [Zr/Fe] and [Ba/Fe] versus [Fe/H] for the
program stars.  The three CH stars (HD\,23439A and B and BD\,+5$^{\rm o}$\,3640)
are shown with filled-in squares.  The error bars for [Zr/Fe]
are the quadratic sum of the standard deviations of the
Zr and Fe abundances from individual lines.  Likewise for [Ba/Fe].
\label{Zr.Ba}}

\figcaption{Plot of [Nd/Fe] versus [Fe/H] for the program stars.
The two CH stars (HD\,23439A and BD\,+5$^{\rm o}$\,3640) with Nd abundance
determinations are shown with filled-in squares.  The Nd
abundances are based on a single \nion{Nd}{ii} line so a fixed error bar
is shown for [Nd/Fe].
\label{Nd}}

\figcaption{A \nion{Zr}{i} line and a \nion{V}{i} line in the two CH stars
HD\,23439A and B and in a comparison star (HD\,103095).  The \nion{Zr}{i}
and \nion{V}{i} lines have similar excitation potentials, 0.00 and 1.05\,eV,
respectively, so their relative strengths are set primarily
by the relative Zr and V abundances.  The much greater strength
of the \nion{Zr}{i} line relative to the \nion{V}{i} line in HD\,23439A and B
compared with HD\,103095 reveals the enhancement of Zr in
HD\,23439A and B.
\label{CH}}
 
\figcaption{ a) The runs of the ``heavy'' neutron-capture elements
Eu, Nd, and Ba as a function of [Fe/H].  The [el/Fe] for Eu and Ba,
which are primarily $r$- and $s$-process elements, respectively, are
schematic representations of the observed [el/Fe], while that for
Nd is predicted on the basis of the behavior of Eu and Ba.  See text
for details.
             b) The predicted runs of the ``light'' neutron-capture
elements Rb and Y as a function of [Fe/H].  These predicted abundances
for Rb and Y, which also rest on the behavior of Eu and Ba,
include allowance for the evolution of [hs/ls] with [Fe/H].  Two
predictions are shown for Rb.  See text for details.
\label{calc.abun}}


\clearpage
\begin{figure}
\psfig{file=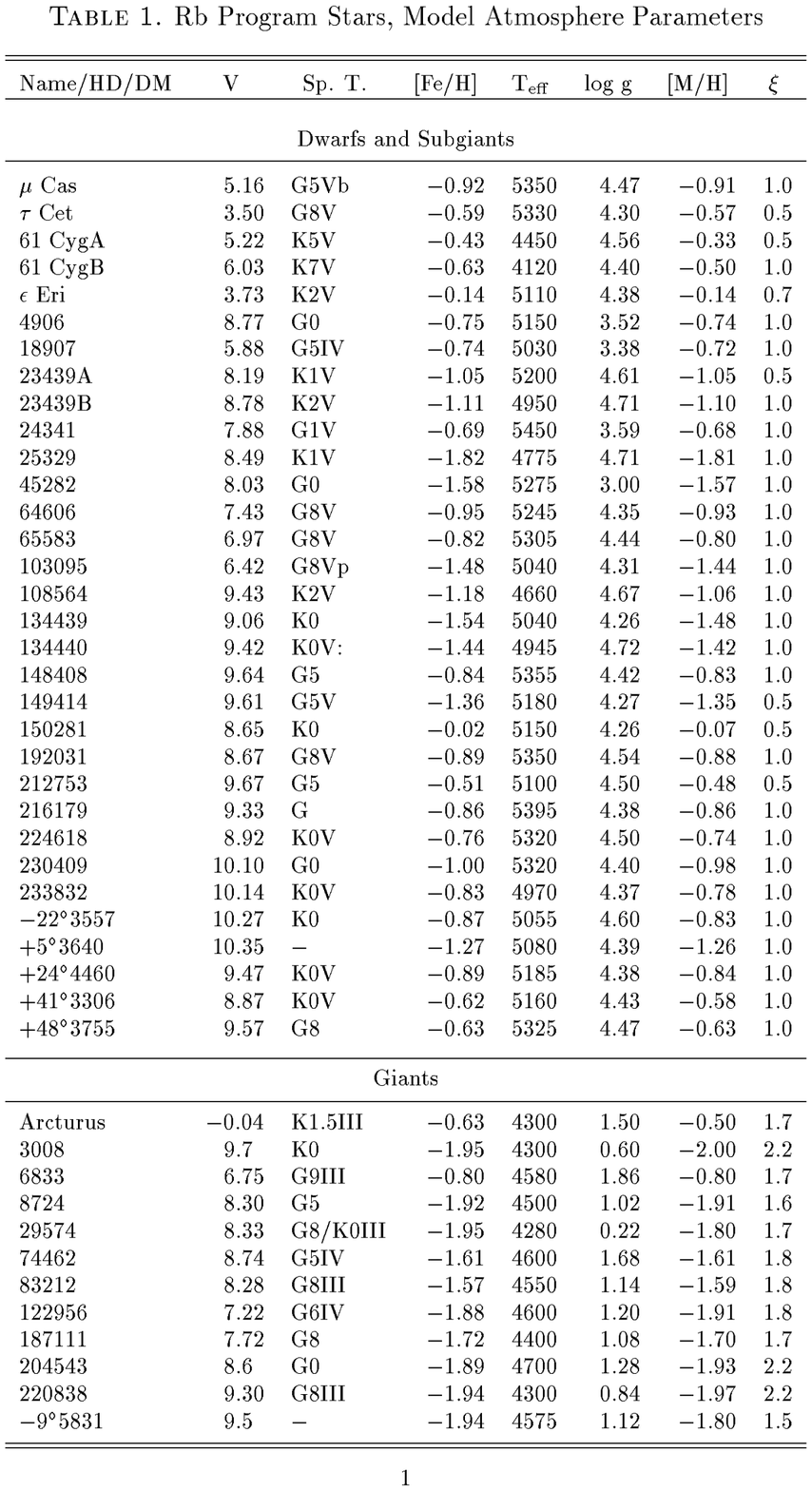,height=9.5in,width=7.5in}
\end{figure}

\clearpage
\begin{figure}
\psfig{file=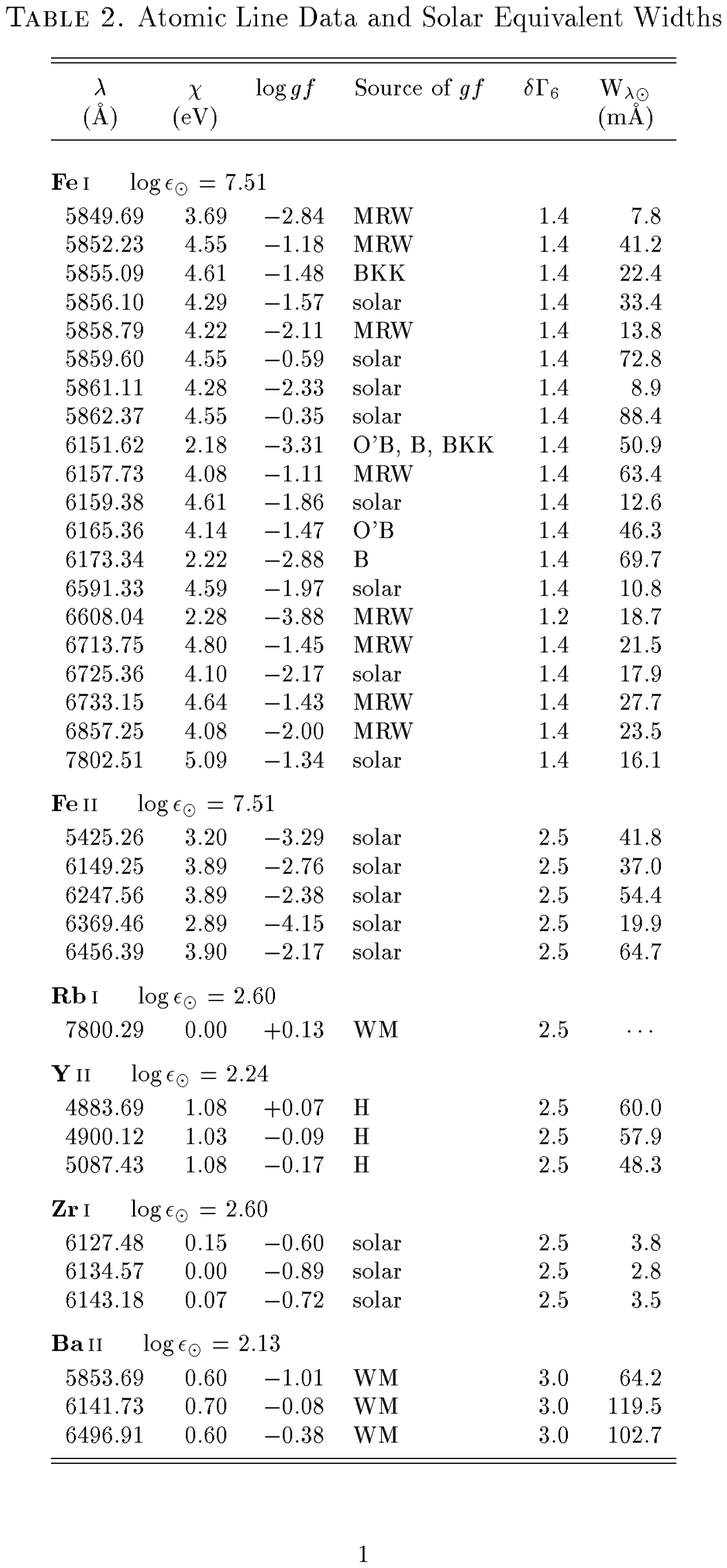,height=10.0in,width=7.5in}
\end{figure}

\clearpage
\begin{figure}
\psfig{file=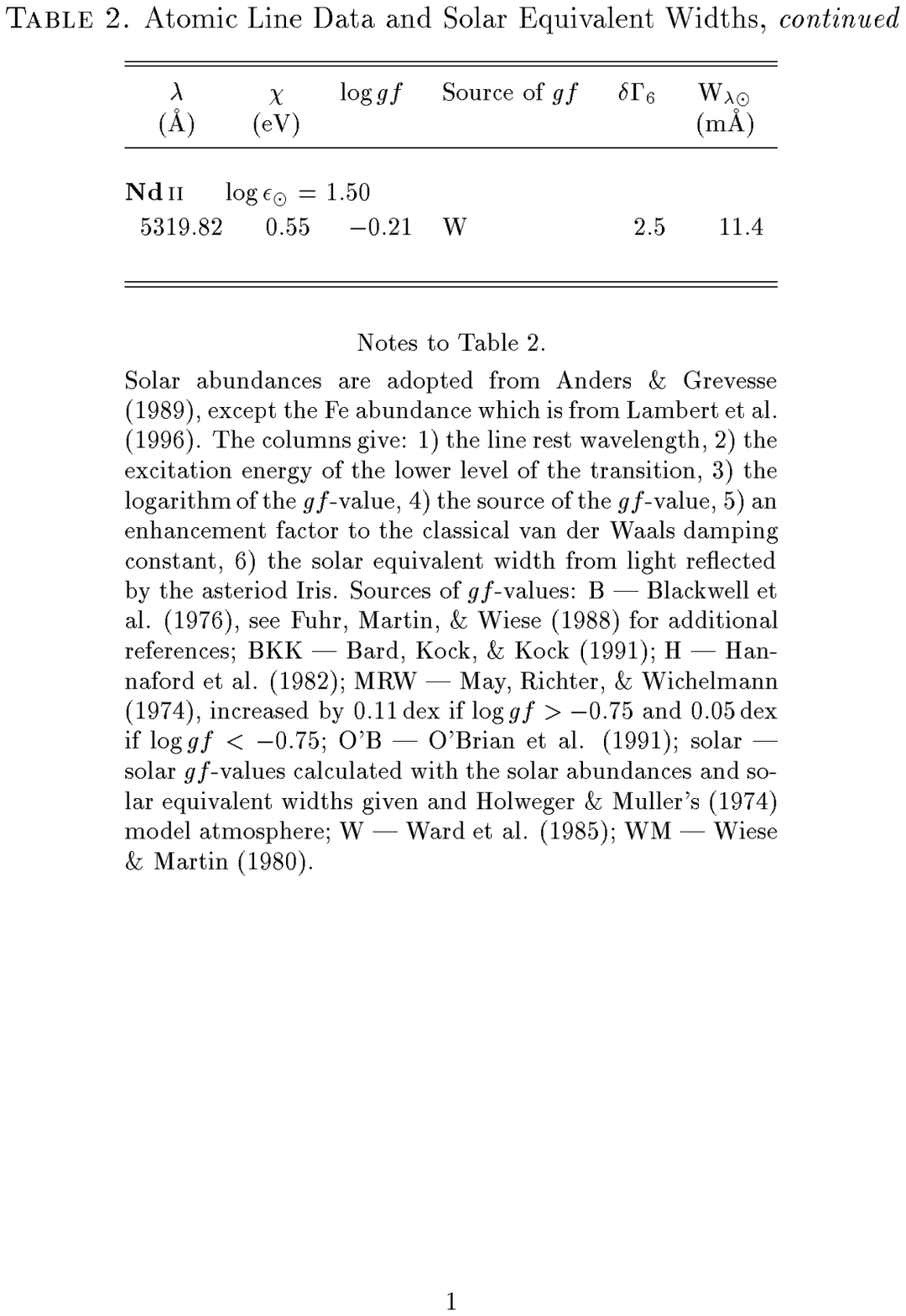,height=10.0in,width=7.5in}
\end{figure}

\clearpage
\begin{figure}
\psfig{file=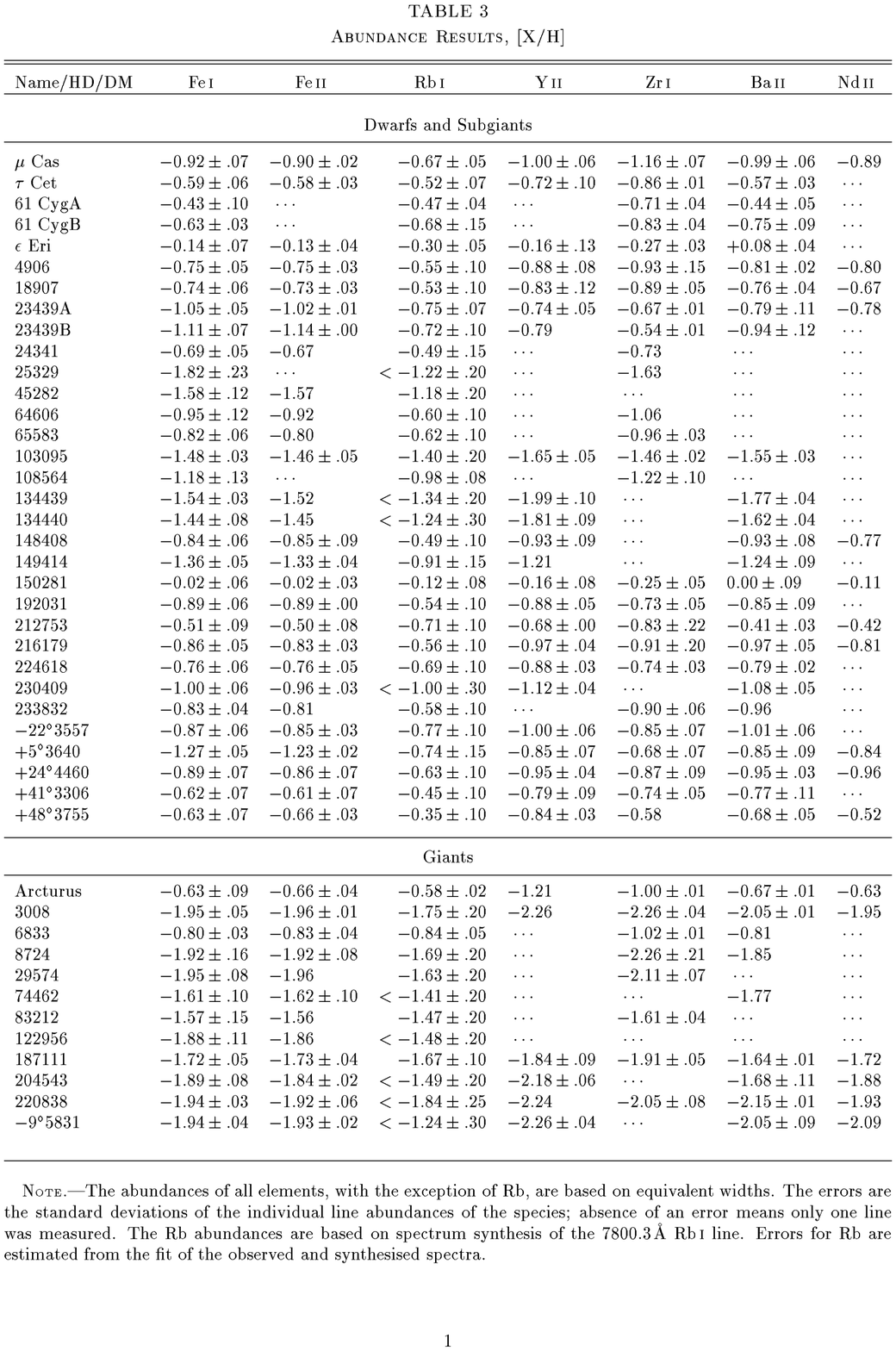,height=10.0in,width=7.5in}
\end{figure}

\clearpage
\begin{figure}
\psfig{file=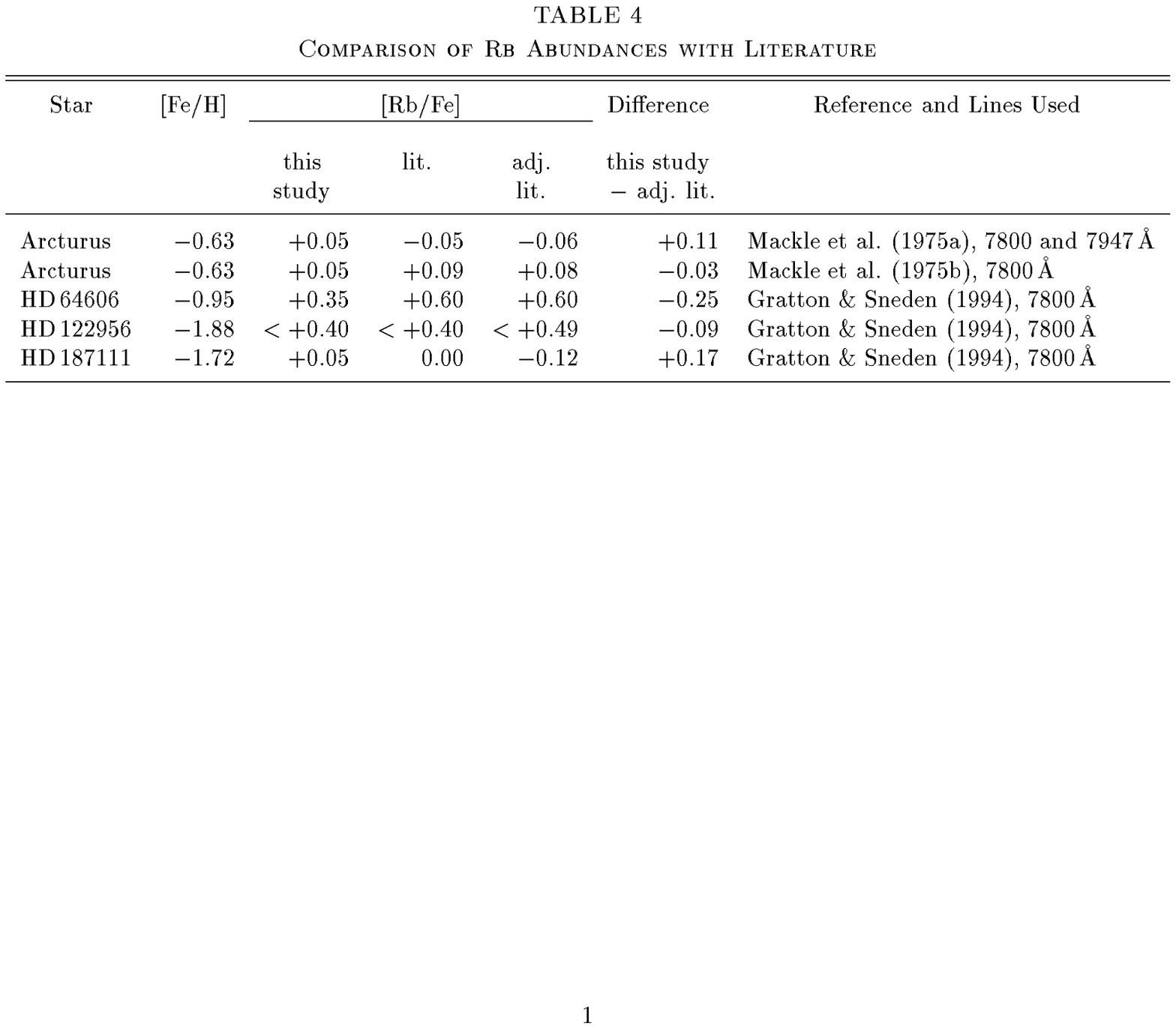,height=10.0in,width=7.5in}
\end{figure}

\clearpage
\begin{figure}
\psfig{file=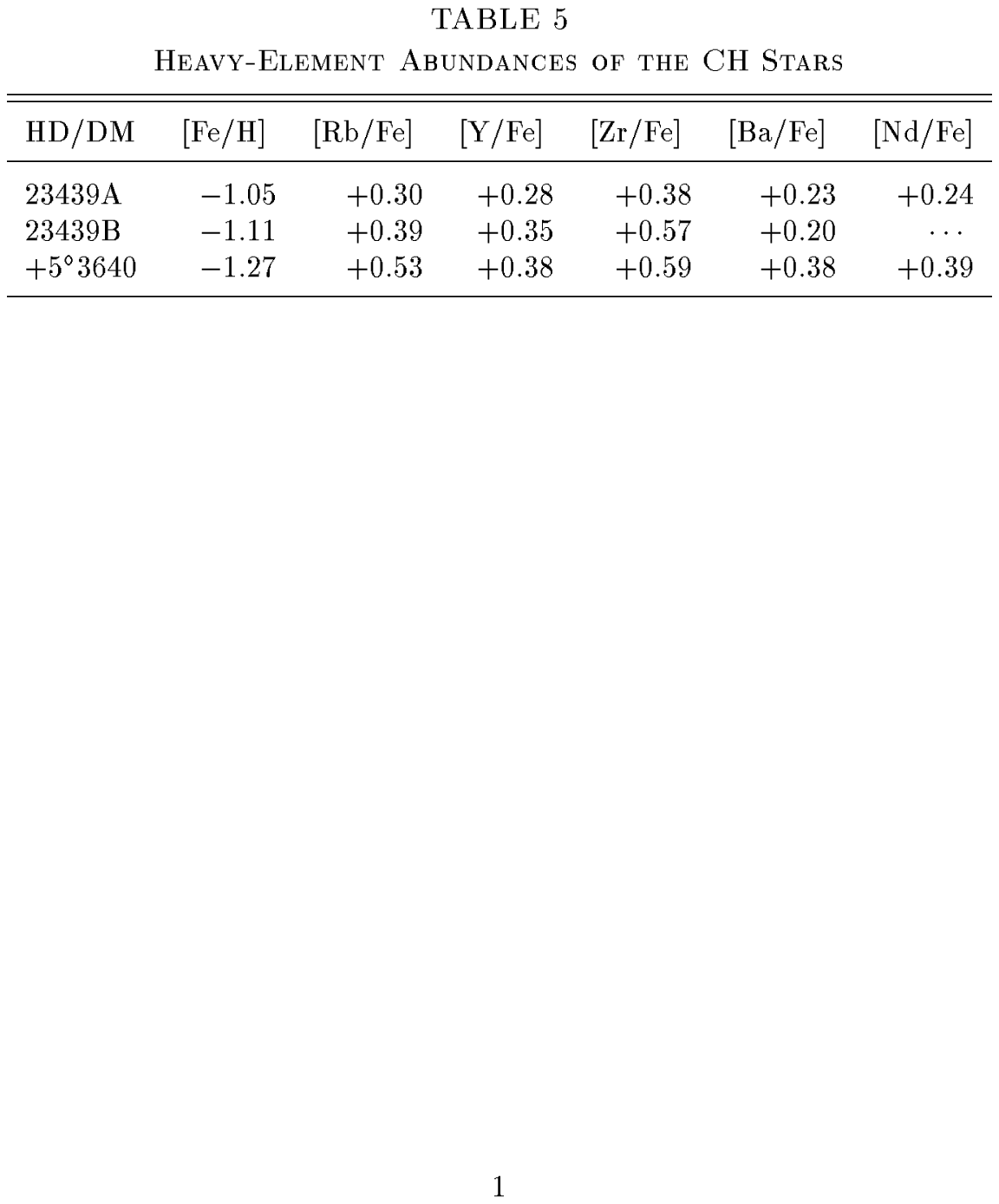,height=10.0in,width=7.5in}
\end{figure}


\clearpage
\begin{figure}
\psfig{file=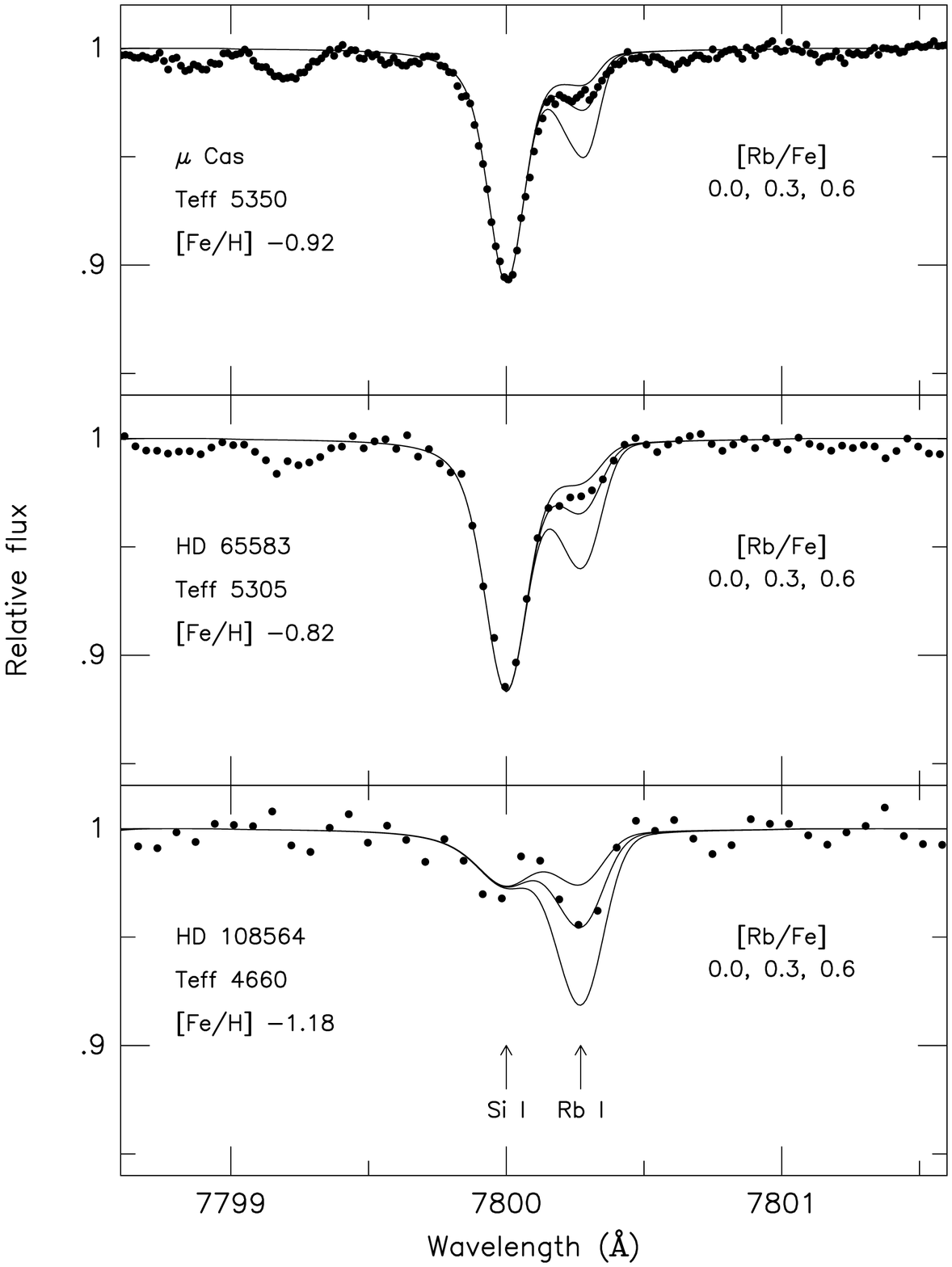,height=8.5in,width=6.5in}
\caption{      ACTUAL FIGURE 1}
\end{figure}

\clearpage
\begin{figure}
\psfig{file=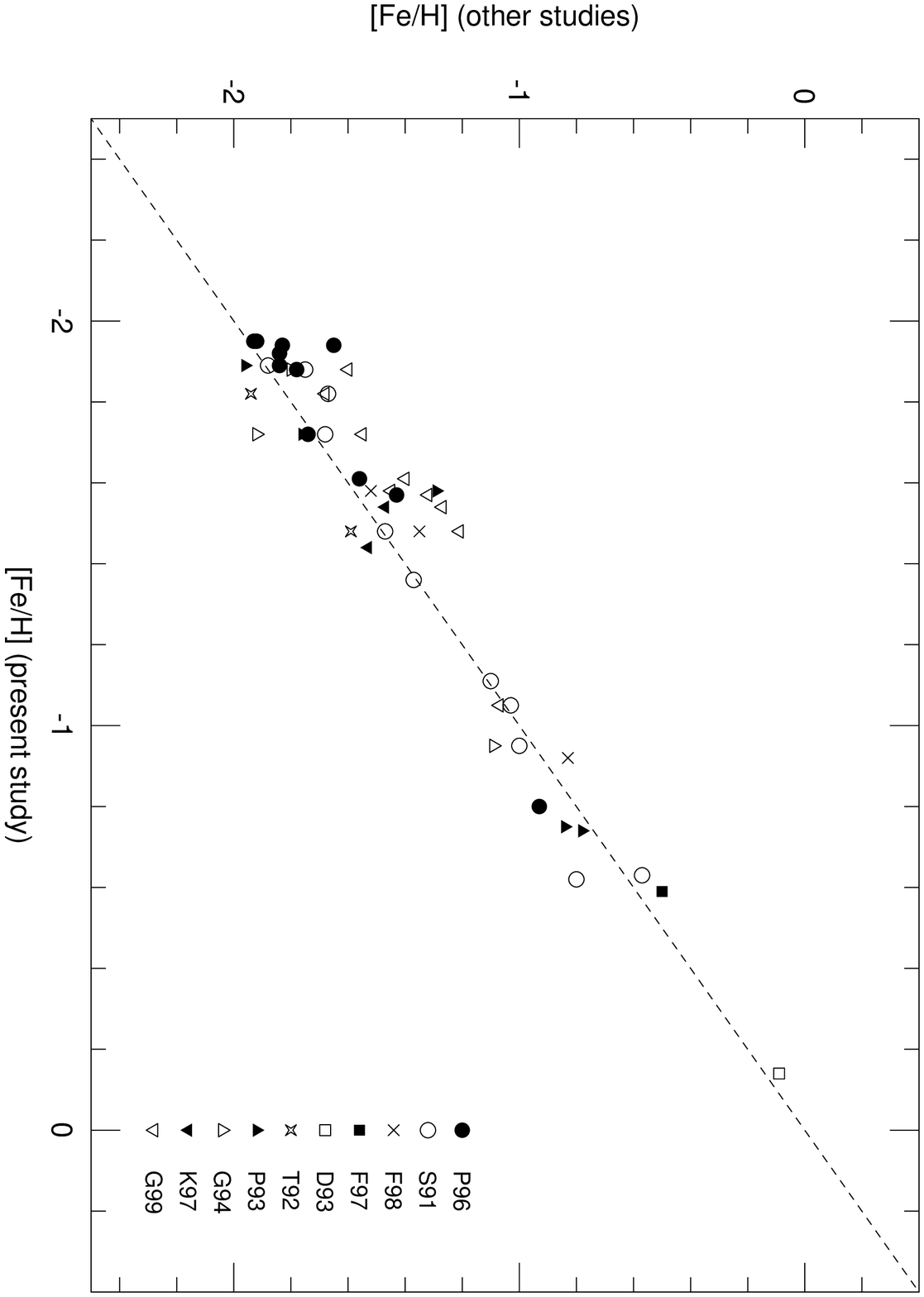,height=8.5in,width=6.5in}
\caption{     ACTUAL FIGURE 2} 
\end {figure}

\clearpage
\begin{figure}
\psfig{file=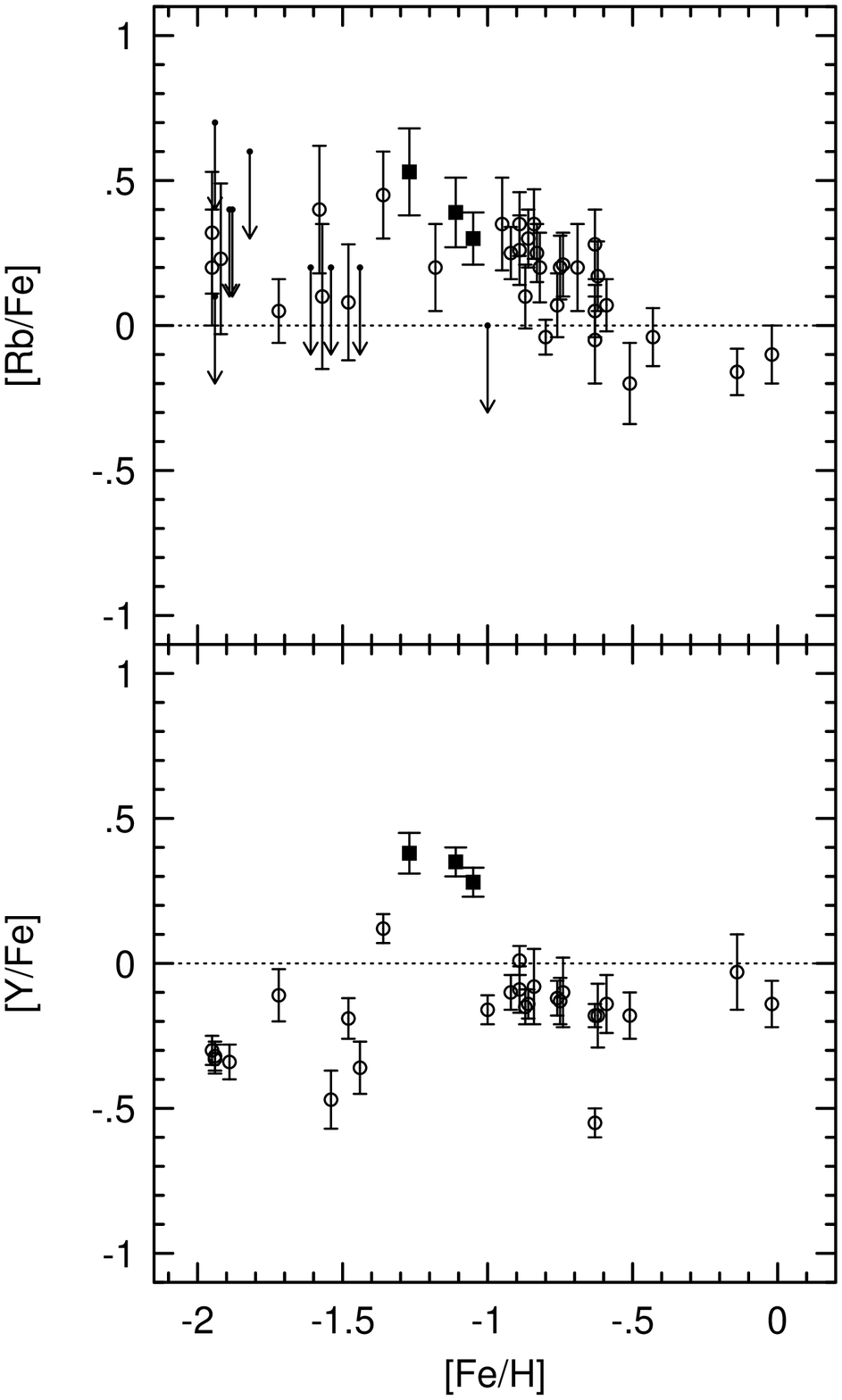,height=8.5in,width=6.5in}
\caption{     ACTUAL FIGURE 3}
\end {figure}

\clearpage
\begin{figure}
\psfig{file=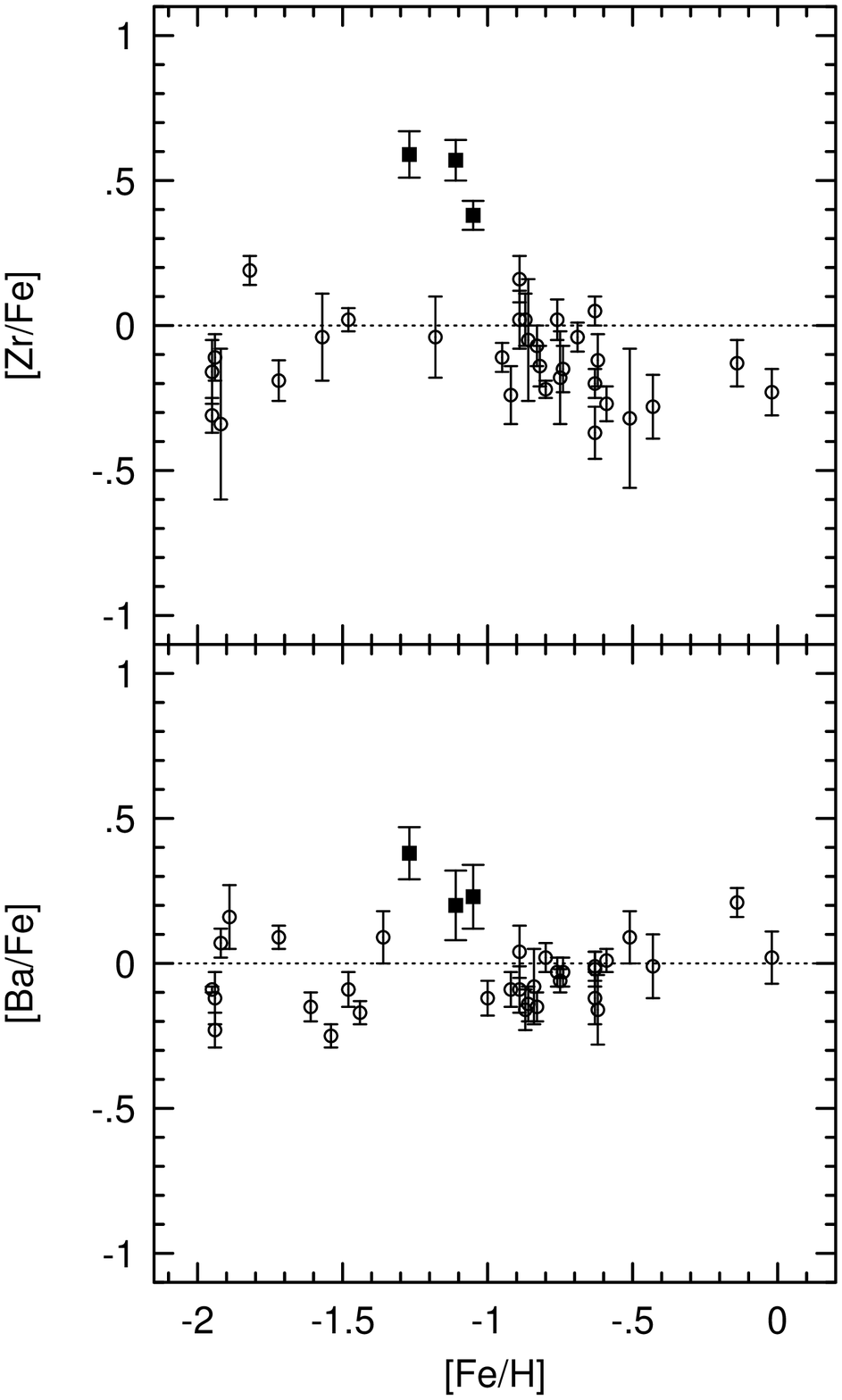,height=8.5in,width=6.5in}
\caption{     ACTUAL FIGURE 4}
\end{figure}

\clearpage
\begin{figure}
\psfig{file=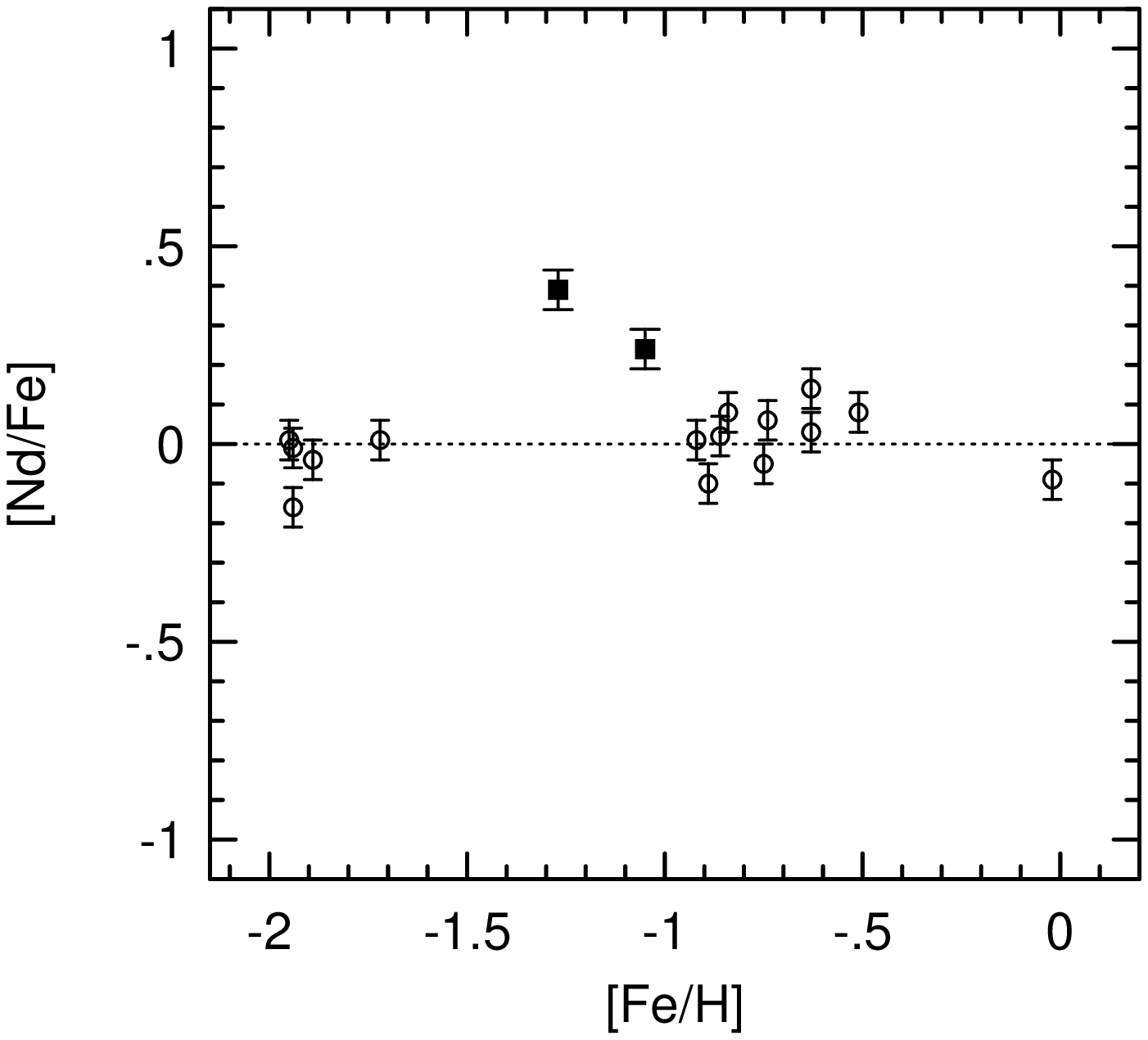,height=8.5in,width=6.5in}
\caption{     ACTUAL FIGURE 5}
\end{figure}

\clearpage
\begin{figure}
\psfig{file=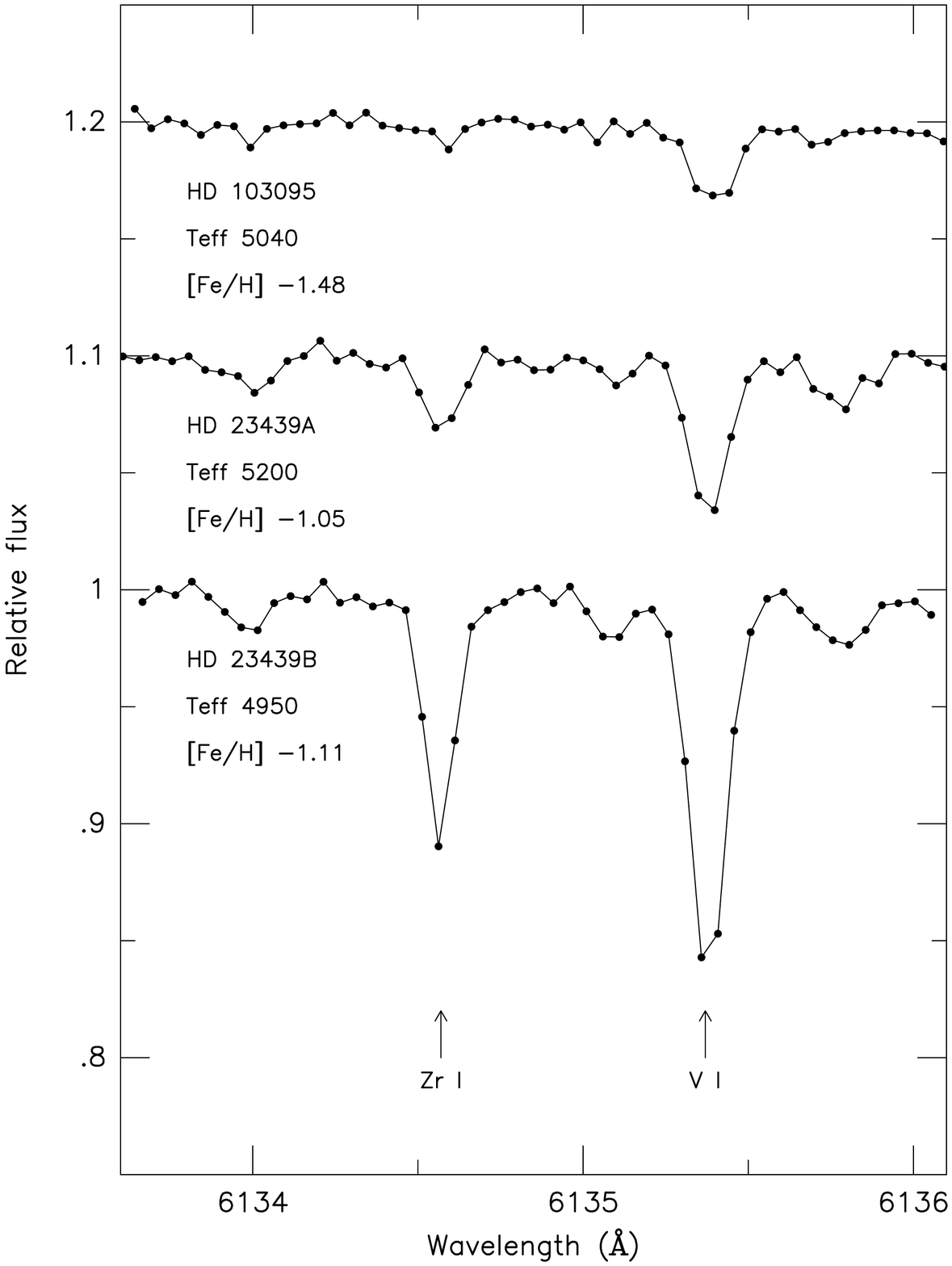,height=8.5in,width=6.5in}
\caption{     ACTUAL FIGURE 6}
\end{figure}

\clearpage
\begin{figure}
\psfig{file=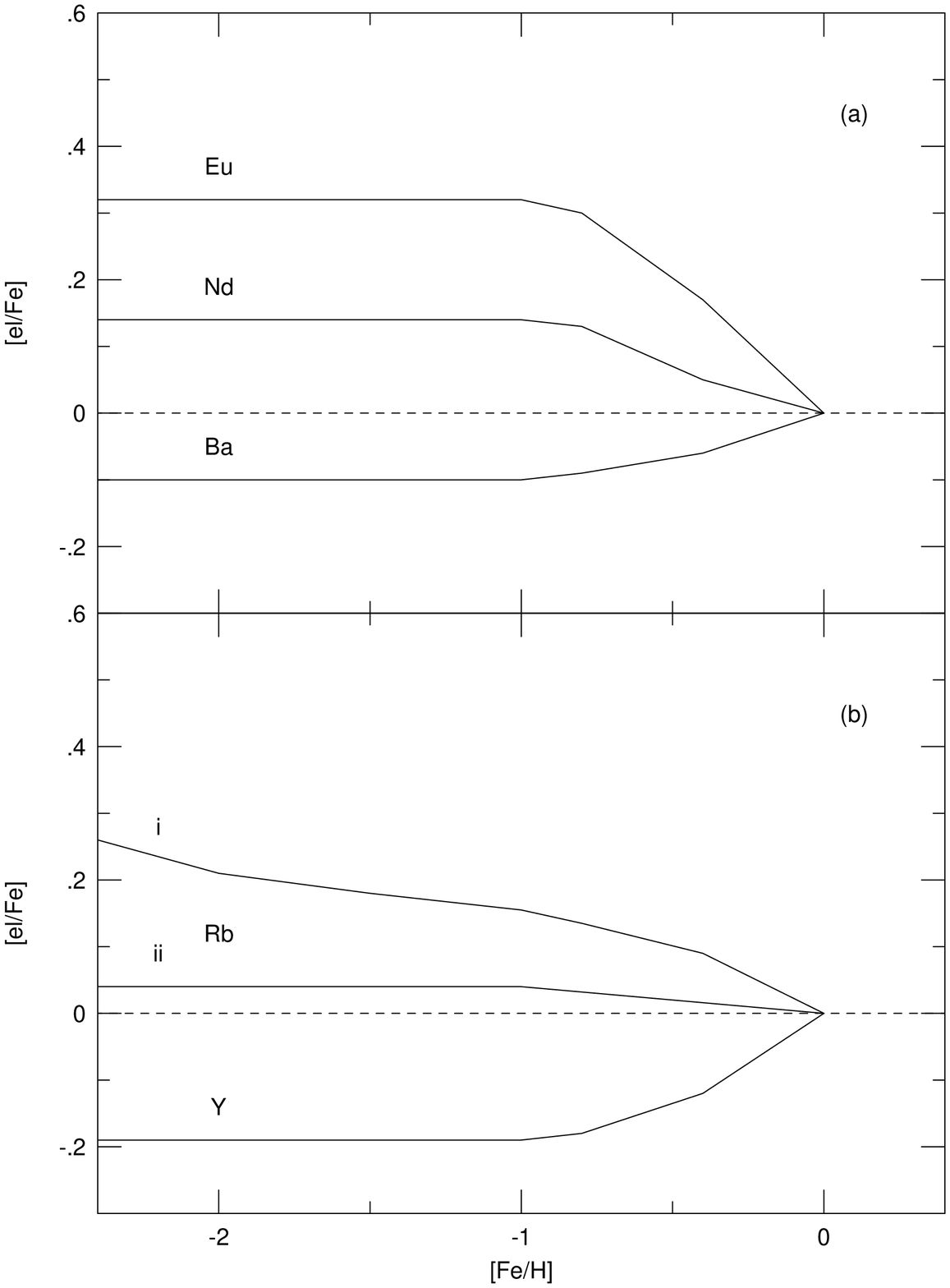,height=8.5in,width=6.5in}
\caption{     ACTUAL FIGURE 7}
\end{figure}


\clearpage

\begin{thebibliography}{}

\bibitem[]{A73}
Allen, C. W. 1973, Astrophysical Quantities (3rd ed.; London: Athlone Press)

\bibitem[]{AGL99}
Allende Prieto, C., Garc\'{\i}a Lop\'ez, R. J., Lambert, D. L., \&
Gustafsson, B. 1999, \apj, submitted

\bibitem[]{AAM96a}
Alonso, A., Arribas, S., \& Mart\'{\i}nez-Roger, C. 1996a, \aap, 313, 873

\bibitem[]{AAM96b}
Alonso, A., Arribas, D., \& Mart\'{\i}nez-Roger, C. 1996b, \aaps, 117, 227

\bibitem[]{AG89}
Anders, E. \& Grevesse, N. 1989, \gca, 53, 197

\bibitem[]{BKK91}
Bard, A., Kock, A., \& Kock, M. 1991, \aap, 248, 315

\bibitem[]{BM89}
Beer, H. \& Macklin, R. L. 1989, \apj, 339, 962

\bibitem[]{BGH81}
Bi\'emont, E., Grevesse, N., Hannaford, P., \& Lowe, R. M. 1981, \apj, 248, 867

\bibitem[]{BIP76}
Blackwell, D. E., Ibbetson, P. A., Petford, A. D., \& Willis, R. B.
1976, \mnras, 177, 219

\bibitem[]{BTL83}
Brown, J. A., Tomkin, J., \& Lambert, D. L. 1983, \apjl, 265L, 93

\bibitem[]{CRB94}
Carlsson, M., Rutten, R. J., Bruls, J. H. M. J., \&
Shchukina, N. G. 1994, \aap, 288, 860

\bibitem[]{CLL94}
Carney, B. W., Latham, D. W., Laird, J. B., Aguilar, L. A. 1994, \aj, 107, 2240

\bibitem[]{C98}
Cowan, J. J. 1998, private communication

\bibitem[]{DS93}
Drake, J. J. \& Smith, G. 1993, \apj, 412, 797

\bibitem[]{FG98}
Feltzing, S. \& Gustafsson, B. 1998, \aaps, 129, 237

\bibitem[]{FM97}
Flynn, C. \& Morell, O. 1997, \mnras, 286, 617

\bibitem[]{FMW88}
Fuhr, J. R., Martin, G. A., \& Wiese, W. L. 1988, J. Phys. Chem. Ref. Data,
17, Suppl. No. 4

\bibitem[]{F98}
Fuhrmann, K. 1998, \aap, 338, 161

\bibitem[]{GJ79}
Gliese, W. \& Jahreiss, H. 1979, \aaps, 38, 423

\bibitem[]{GCB99}
Gratton, R. G., Carretta, E., Bragalia, A., \& Sneden, C. 1999,
in preparation

\bibitem[]{GS94}
Gratton, R. G. \& Sneden, C. 1994, \aap, 287, 927

\bibitem[]{GBE75}
Gustafsson, B., Bell, R. A., Eriksson, K., \& Nordlund, \AA. 1975, \aap, 42, 407

\bibitem[]{HHM86}
Hackett, P. A., Humphries, M. R., Mitchell, S. A., \& Rayner, D. M. 1986,
\jcp, 85, 3194

\bibitem[]{HLG82}
Hannaford, P., Lowe, R. M., Grevesse, N., Bi\'emont, E., \& Whaling, W.
1982, \apj, 261, 736

\bibitem[]{HJ82}
Hoffleit, D. \& Jaschek, C. 1982, The Bright Star Catalogue (Yale Univ.
Observatory, New Haven)

\bibitem[]{HM74}
Holweger, H. \& M\"uller, E. A. 1974, \solphys, 39, 19

\bibitem[]{KBW89}
K\"appeler, F., Beer, H., \& Wisshak, K. 1989, Rep. Prog. Phys., 52, 945

\bibitem[]{K97}
King, J. R. 1997, \aj, 113, 2302

\bibitem[]{KFB84}
Kurucz, R. L., Furenlid, I., Brault, J., \& Testerman, L. 1984,
Solar Flux Atlas from 296 to 1300 nm, Cambridge, MA: Harvard Univ.

\bibitem[]{L89}
Lambert, D. L. 1989, in: Cosmic Abundances of Matter; Proceedings
of the AIP Conference, Minneapolis, MN, Sept. 7-9, 1988
(A90-31833 13-90). New York, American Institute of Physics, 1989, p.168

\bibitem[]{LHL96}
Lambert, D. L., Heath, J. E., Lemke, M., \& Drake, J. 1996, \apjs, 103, 183

\bibitem[]{LL76}
Lambert, D. L. \& Luck, R. E. 1976, Observatory, 96, 100

\bibitem[]{LSB95}
Lambert, D. L., Smith, V. V., Busso, M., Gallino, R., \& Straniero, O.
1995, \apj, 450, 302

\bibitem[]{LMC88}
Latham, D. W., Mazeh, T., Carney, B. W., McCrosky, R. E., Stefanik, R. P.,
\& Davis, R. J. 1988, \aj, 96, 567

\bibitem[]{MGG75b}
M\"ackle, R., Griffin, R., Griffin, R., \& Holweger, H. 1975b, \aaps, 19, 303

\bibitem[]{MHG75a}
M\"ackle, R., Holweger, H., Griffin, R., \& Griffin, R. 1975a, \aap, 38, 239

\bibitem[]{MRW74}
May, M., Richter, J., \& Wichelmann, J. 1974, \aaps, 18, 405

\bibitem[]{M97}
McWilliam, A. 1997, \araa, 35, 503

\bibitem[]{MMH66}
Moore, C. E., Minnaert, M. G. J., \& Houtgast, J.
1966, The Solar Spectrum 2935{\AA} to 8770{\AA}, NBS Mono. 61,
Washington, U.S. Gov.

\bibitem[]{NHS97}
Nissen, P. E., H{\o}g, E., \& Schuster, W. J. 1997,
Proceedings of the ESA Symposium `Hipparcos - Venice '97', 13-16 May, Venice,
Italy, ESA SP-402 (July 1997), p. 225

\bibitem[]{NS91}
Nissen, P. E. \& Schuster, W. J. 1991, \aap, 251, 457

\bibitem[]{OWL91}
O'Brian, T. R., Wickliffe, M. E., Lawler, J. E., Whaling, W.,
\& Brault, J. W. 1991, J. Opt. Soc. A., B8, 1185

\bibitem[]{PSB93}
Pilachowski, C. A., Sneden, C. \& Booth, J. 1993, \apj, 407, 699

\bibitem[]{PSK96}
Pilachowski, C. A., Sneden, S., \& Kraft, R. P. 1996, \aj, 111, 1689

\bibitem[]{SN88}
Schuster, W. J. \& Nissen, P. E. 1988, \aaps, 73, 225

\bibitem[]{SN89a}
Schuster, W. J. \& Nissen, P. E. 1989a, \aap, 222, 69

\bibitem[]{SN89b}
Schuster, W. J. \& Nissen, P. E. 1989b, \aap, 221, 65

\bibitem[]{S97}
Smith, V. V. 1997, in Synthesis of the Elements in Stars: Forty Years of
Progress; Wallerstein, G., Iben, I. Jr., 
Parker, P., Boesgaard, A. M., Hale, G. M., Champagne, A. E., Barnes, C. A., 
K{\"a}ppeler, F., Smith, V. V., Hoffman, R. D., Timmes, F. X., Sneden, C., 
Boyd, R. N., Meyer, B. S., and Lambert, D. L., Rev. Mod. Phys., 69, 995

\bibitem[]{S73}
Sneden, C. 1973, PhD thesis, University of Texas at Austin

\bibitem[]{SC88}
Sneden, C. \& Crocker, D. A. 1988, \apj, 335, 406

\bibitem[]{SGC91}
Sneden, C., Gratton, R. G., \& Crocker, D. A. 1991, \aap, 246, 354

\bibitem[]{SP83}
Sneden, C., and Parthasarathy, M., 1983, \apj, 267, 757

\bibitem[]{TEL97}
Tomkin, J., Edvardsson, B., Lambert, D. L., \& Gustafsson, B. 1997,
\aap, 327, 587

\bibitem[]{TL83}
Tomkin, J. \& Lambert, D. L. 1983, \apj, 273, 722

\bibitem[]{TLL92}
Tomkin, J., Lemke, M., Lambert, D. L. \& Sneden, C. 1992, \aj, 104, 1568

\bibitem[]{T81}
Truran, J. W. 1981, \aap, 97, 391

\bibitem[]{TMS95}
Tull, R. G., MacQueen, P. J., Sneden, S., \& Lambert, D. L. 1995,
\pasp, 107, 251

\bibitem[]{WVA85}
Ward, L., Vogel, O., Arnesen, A., Hallin, R., \& W\"annstr\"om, A. 1985,
Physica Scripta, 31, 161

\bibitem[]{WST89}
Wheeler, J. C., Sneden, C., \& Truran, J. W. Jr. 1989, \araa, 27, 279

\bibitem[]{WM80}
Wiese, W. L. \& Martin, G. A. 1980, ``Wavelengths and Transition
Probabilities for Atoms and Atomic Ions'', NSRDS-NBS 68
(U.S. Government Printing Office, Washington,D.C.)

\bibitem[]{ZM91}
Zhao, G. \& Magain, P. 1991, \aap, 244, 425

\end{thebibliography}
\end{document}